\documentclass[useAMS,usenatbib]{mn2e}
\usepackage{graphicx}
\usepackage{BibDef}
\usepackage{amssymb}
\usepackage[english]{babel}
\usepackage{natbib}
\usepackage{color}
\usepackage{rotating}
\usepackage{pdflscape}

\title[Relativistic Fe K$\alpha$ line study in Seyfert 1 galaxies observed with \textit{SUZAKU}]{Relativistic Fe K$\alpha$ line study in Seyfert 1 galaxies observed with \textit{SUZAKU} }
\author[G. Mantovani, K. Nandra, G. Ponti]{G. Mantovani\thanks{E-mail: gmantova@mpe.mpg.de}, K. Nandra, G. Ponti
\\
Max-Planck-Institut f\"{u}r Extraterrestrische Physik, Giessenbachstrasse 1, D-85748 Garching bei M\"{u}nchen, Germany}
\begin{document}

\date{Accepted 2016 March 9. Received 2016 March 9; in original form 2015 June 3}

\pagerange{\pageref{firstpage}--\pageref{lastpage}} \pubyear{xxxx}

\maketitle

\label{firstpage}

\begin{abstract}
We present an analysis of a sample of Seyfert 1 galaxies observed with \textit{Suzaku}. The aim of this work is to examine critically the evidence for a relativistic Fe K$\alpha$ line in the X-ray spectra of these AGN. The sample was compiled from those sources in which a relativistic component was missing in at least one \textit{XMM-Newton} observation. We analysed the \textit{Suzaku} spectra of these objects in order to have more constraints on the high energy emission, including the Compton reflection hump. The results show that the relativistic Fe K$\alpha$ line is detected (at $>95\%$ confidence) in all sources observed with high signal-to-noise ratio (e.g. where the counts in the 5-7 keV energy band are $\gtrsim$ 4 $\times$ 10$^4$). This is in agreement with the idea that relativistic lines are a ubiquitous feature in the spectra of Seyfert galaxies, but are often difficult to detect without very high quality data. We also investigate the relation between the Fe K$\alpha$ line and the reflection continuum at high energies. For most of the sample, the strength of the reflection component is consistent with that of the line. There are exceptions in both senses, however i.e. where the reflection continuum is strong but with weak line emission, and vice versa. These observations present a challenge for standard reflection models. 
\end{abstract}


\section{Introduction}

According to standard models, the X-ray emission of Active Galactic Nuclei (AGN) originates in the innermost regions of the accretion flow, via inverse Compton scattering of the disc thermal photons in a hot, optically thin plasma (the so called corona) located above the accretion disc (e.g., \citealt{Shapiro+76}; \citealt{Sunyaev+80}; \citealt{Haardt+93}; \citealt{HaardtM+93}; \citealt{Haardt+94}). 

This primary X-ray emission interacts with the optically thick matter of the accretion disc and the molecular torus, producing a reflection spectrum \citep{George+91}. This spectrum is characterised by a series of fluorescence lines, the most prominent, due to a combination of fluorescence yield and cosmic abundance, is the Fe K$\alpha$ line at 6.4 keV (e.g., \citealt{Matt+97}; \citealt{nandra+94}). The reflection spectrum extends also to the hard X-ray band, where emission is produced as a consequence of Compton-back scattering of the high energy photons (e.g., \citealt{Pounds+90}; \citealt{George+91}) from the primary continuum. This reflection continuum typically peaks at $\sim$20-30 keV producing a prominent broad feature, known as the Compton hump. When the reflection arises from the inner regions of the accretion disc, it will be affected by the strong gravitational field of the black hole. The result is an Fe K$\alpha$ emission line whose shape is modified into a skewed and asymmetric profile (\citealt{Fabian+89}, \citeyear{Fabian+02}, \citeyear{Fabian+09}, \citealt{Tanaka+95}). Since the relativistic lines originate within few gravitational radii from the central object, the analysis of these features represents a powerful probe of the innermost region of the AGN. 

Analysis of the Fe line complex (i.e. \citealt{Yaqoob+04}; \citealt{Nandra+07}; \citealt{Calle+10}; \citealt{Patrick+12}) has revealed the presence of a ubiquitous narrow  component in the spectra of the brightest AGN observed with \textit{XMM-Newton} and \textit{SUZAKU}. Assuming a standard accretion disc, expected in high efficiency systems such as bright Seyfert galaxies, broad Fe K$\alpha$ lines should also be detected in the X-ray spectra. However, in samples of objects, where \textit{XMM-Newton} and \textit{SUZAKU} observations were analysed, some sources and/or observations were missing this component (30$\%$ of the sources in \citealt{Nandra+07}, 60$\%$ in \citealt{Calle+10}, 50$\%$ in \citealt{Patrick+12} and 20$\%$ in \citealt{Walton+13}). The reason why this component is not detected in some cases is still unclear. Several hypotheses have been proposed to explain this puzzle. Looking at the samples compiled with \textit{XMM-Newton} observations, the overall picture shows that low signal-to-noise ratio could be one of the main reasons for the absence of a detection of the relativistic line in some observations or sources (e.g., \citealt{Guainazzi+06}, \citealt{Nandra+07}, \citealt{Calle+10}). Another possible explanation is that in some sources the disk is truncated, so that the inner radius is large and the relativistic effects relatively small. \citet{Bhayani+11} have on the other hand suggested that strong relativistic effects, disk ionization and/or high disk inclination can explain the apparent lack of relativistic signatures, due to the difficulty of disentangling very broad features from the continuum. 

The aim of this work is to investigate this phenomenon by analyzing \textit{SUZAKU} observations of sources previously observed with \textit{XMM-Newton}, where a relativistic Fe K$\alpha$ line is missing. The high throughput of \textit{SUZAKU} around the Fe K$\alpha$ line allows an independent check of the \textit{XMM-Newton} results, while the wide energy range allows an investigation of the Compton hump at high energies, for comparison with the emission line properties. 

\section{Sample}

The objects for our analysis were chosen from the work of \citet{Nandra+07}. They compiled their sample from the observations available in the \textit{XMM-Newton} public archive, cross-correlated with the \citet{Veron+01} AGN catalogue. The analysis was restricted to nearby objects, with a cutoff on the redshift of $z<$ 0.05. Seyfert 2 galaxies were excluded, as were radio loud objects and central cluster galaxies (see \citet{Nandra+07} for more details). From these objects we selected those in which a relativistic Fe K$\alpha$ line was not detected. If a source has multiple observations and one of these did not show a relativistic component, the object was still included in our analysis. The sources satisfying these criteria and observed by \textit{SUZAKU} are reported in Table \ref{Sample}. The \textit{SUZAKU} data give better constraints at the high energies ($>10$keV) and hence better modelling of the primary continuum and the Compton hump. Two sources, Mrk 6 and HE 1143-1810, were excluded in the following analysis because no \textit{SUZAKU} observations are available. In Table \ref{Sample} we report the continuum flux in the Fe K band (5-7 keV) of each source. When multiple observations are available, the mean flux is quoted. We also report the number of observations analysed and the total exposure time in ks. 

\begin{table}
\centering
\caption{AGN sample. These consist of objects observed by Suzaku which are missing a relativistic Fe K$\alpha$ line in the \textit{XMM-Newton} observations (\citealt{Nandra+07}). The flux is in units of 10$^{-11}$ erg cm$^{-2}$ s$^{-1}$. \dag{Flux in the 2-10 keV band from \citealt{Nandra+07}.}}
\begin{minipage}{140mm}
\label{Sample}
\begin{tabular}{@{}cccc@{}}
\hline
Object & Flux 5-7 keV & $\#$Suzaku Obs& Exposure (ks)\\
\hline
NGC 2110 & 3.4 & 2 & 205.300\\
NGC 5506 & 2.7 & 3 & 158.455\\
IC 4329A & 2.6 & 6 & 248.820\\
MCG +8-11-11 & 1.6 & 1 & 98.750\\
NGC 7213 & 0.57 & 1 & 90.750\\
MRK 110 & 0.48 & 1 & 90.900\\
NGC 7469 & 0.49 & 1 & 112.100\\
NGC 5548 & 0.46 & 7 & 209.435\\
MRK 590 & 0.18 & 2 & 102.520\\ 
HE 1143-1810 & 2.83\dag & 0 & -\\
MRK 6 & 1.43\dag & 0 & -\\ \hline
\end{tabular}
\end{minipage}
\end{table}

A special case is that of NGC 2110. This is a narrow-line object in the optical and is viewed through a relatively large absorption screen, which may be patchy (e.g., \citealt{Evans+07}), or ionized (\citealt{Nandra+07}). Different observations with \textit{XMM-Newton}, \textit{Chandra} and \textit{SUZAKU} demonstrated the presence of a relatively stable, full-covering absorber (column density of $\sim$ 3 $\times$ 10$^{22}$ cm$^{-2}$). Together with this, an additional absorber, likely variable in both column density and covering fraction, was required (\citealt{Rivers+14}). The difficulties in modelling the absorber could produce residuals in the Fe band, which are difficult to disentangle from any relativistic Fe K$\alpha$ line. As the latter is the main focus of our current work, we chose to exclude this source from the following spectral analysis.

\section{Data Reduction}
\label{Reduction}

\textit{SUZAKU} X-ray Imaging Spectrometer (XIS; \citealt{Koyama+07}) and Hard X-ray Detector (HXD; \citealt{Kokubun+07}; \citealt{Takahashi+07}) event files for each observation were processed adopting the standard filtering criteria.  We used the calibration files from the 2013-01-10 release and the \textit{FTOOLS} package of lheasoft version 6.13\footnote{http://heasarc.gsfc.nasa.gov/ftools/} (\citealt{Blackburn+95}). 

\subsection{X-ray Imaging Spectrometer}

Source spectra were extracted from the XIS cleaned events files adopting circular regions of 250 arcsec, centered on the source. The background regions were made as large as possible avoiding the source, the calibration regions at the chip corners, and any obvious non-nuclear emission. Response matrices and ancillary response files were produced with \textit{xisrmfgen} and \textit{xissimarfgen}. Data from the XIS0 and XIS3 detectors were used, and the spectra were added together. Finally, the spectra were rebinned in order to have at least 100 counts per bin. We did not include the back-illuminated XIS1 spectra because of the lower effective area at 6 keV and higher background level at high energies compared to the front illuminated detectors XIS0 and XIS3.

\subsection{Hard X-ray Detector}

The PIN data were reduced following the standard procedure, as presented in the \textit{SUZAKU} ABC guide\footnote{http://heasarc.gsfc.nasa.gov/docs/suzaku/analysis/abc/}. The spectra were corrected for dead time and the exposure of the background spectra was increased by a factor of 10, as required. The sum of the "tuned" non X-ray background (NXB) spectra and simulations of the Cosmic X-ray Background (CXB) was adopted as PIN background in the spectral analysis. The CXB is simulated from a typical model provided by the \textit{SUZAKU} team \citep{Boldt+87}. It is known that the detected spectrum can be different from that model, due to spatial fluctuations of the intrinsic CXB. These fluctuations are known to be of 10$\%$ from on scales of 1 sq degree \citep{Barcons+2000}. In those cases where the PIN background has important uncertainties and the source flux is relatively low, the reliability of the PIN data can be affected. Since the CXB corresponds only to 5$\%$ of the total background spectrum \citep{Fukazawa+09}, the fluctuations in the total PIN background are dominated by the uncertainties of the NXB, which are measured to be of order  3$\%$. Considering all these issues, we decided to use PIN spectrum only in cases where the source spectrum is more than the 20$\%$ of the background, to avoid being dominate by background uncertainties (see Table \ref{PIN}).

\begin{table}
\centering
\caption{PIN energy ranges used in the analysis.}
\begin{minipage}{140mm}
\label{PIN}
\begin{tabular}{@{}cccc@{}}
\hline
Object & Observation ID & PIN Energy Range\\
\hline
NGC 5506 & 701030010 & 12-60 keV\\
& 701030020 & 12-50 keV\\
& 701030030 & 12-50 keV\\ \\
IC 4329A & 702113010 & 12-55 keV\\
& 702113020 & 12-50 keV\\
& 702113030 & 12-55 keV\\
& 702113040 & 12-70 keV\\
& 702113050 & 12-55 keV\\
& 707025010 & 12-55 keV\\ \\
MCG +8-11-11 & 702112010 & 12-50 keV\\ \\
NGC 7213 & 701029010 & Not Considered\\ \\
MRK 110 & 702124010 & 12-25 keV\\ \\
NGC 7469 & 703028010 & 12-30 keV\\ \\
NGC 5548 & 702042010 & Not Considered\\ 
& 702042020 & Not Considered\\
& 702042040 & 12-25 keV\\
& 702042050 & 12-25 keV\\
& 702042060 & 12-35 keV\\
& 702042070 & 12-30 keV\\
& 702042080 & Not Considered\\ \\
MRK 590 & 705043010 & Not Considered\\
& 705043020 & Not Considered\\ \hline
\end{tabular}
\end{minipage}
\end{table}

After this analysis, the PIN spectra were rebinned to have at least 40 counts per bin after background subtraction. 

\section{Data Analysis}

For the spectral analysis, the \textit{xspec} program (version 12.8.0) was used. All quoted errors correspond to the 90$\%$ confidence level for one interesting parameter ($\Delta\chi^2$ = 2.71), unless otherwise stated. The XIS and PIN spectra were fitted together. To minimise the effects of absorption on the spectral fits, the XIS spectra were analysed in the 3-10 keV band, while the PIN spectra were considered in the range presented in Table \ref{PIN}. We introduced a cross-normalisation constant for both instruments, fixing this to 0.994 the value for XIS0 and XIS3, and to 1.164 for the PIN, as appropriate for data taken at XIS nominal position \citep{Maeda+08}. 

\subsection{ Baseline model}
\label{Baseline}

In general we fitted the spectra systematically with the same models. In a first attempt to model the continuum spectrum, we applied a model consisting of a power law with a high energy cutoff, along with neutral reflection  (\citealt{Magdziarz+95}) and a neutral absorber at the redshift of the source. Specifically, we used the \textit{zwabs*(pexrav)} model (Baseline model). The neutral absorber is required only in the cases of IC 4329A and NGC 5506, with column densities of N$_H$ = 0.4 $\times$ 10$^{22}$ cm$^{-2}$ (\citealt{Mantovani+14}) and N$_H$ $\sim$ 3 $\times$ 10$^{22}$ cm$^{-2}$ (\citealt{Bianchi+03}), respectively. For the reflection, we adopted solar abundance and we assumed an inclination of 60$^{\circ}$. In most cases, the high energy cutoff of the intrinsic primary continuum was fixed to 300 keV. In the case of IC 4329A, we adopted E$_c$ = 180 keV, in accordance with \textit{NuSTAR} and \textit{SUZAKU} observations (\citealt{Brenneman+14}, \citealt{Mantovani+14}). For NGC 5506 we fixed this parameter to E$_c$ = 130 keV, as reported in previous \textit{BeppoSAX} observations (\citealt{Bianchi+03}).

\begin{figure*}
\hspace*{0.5cm} 
\begin{minipage}[b]{0.4\linewidth}
\centering
\includegraphics[width=0.65\textwidth, angle=90]{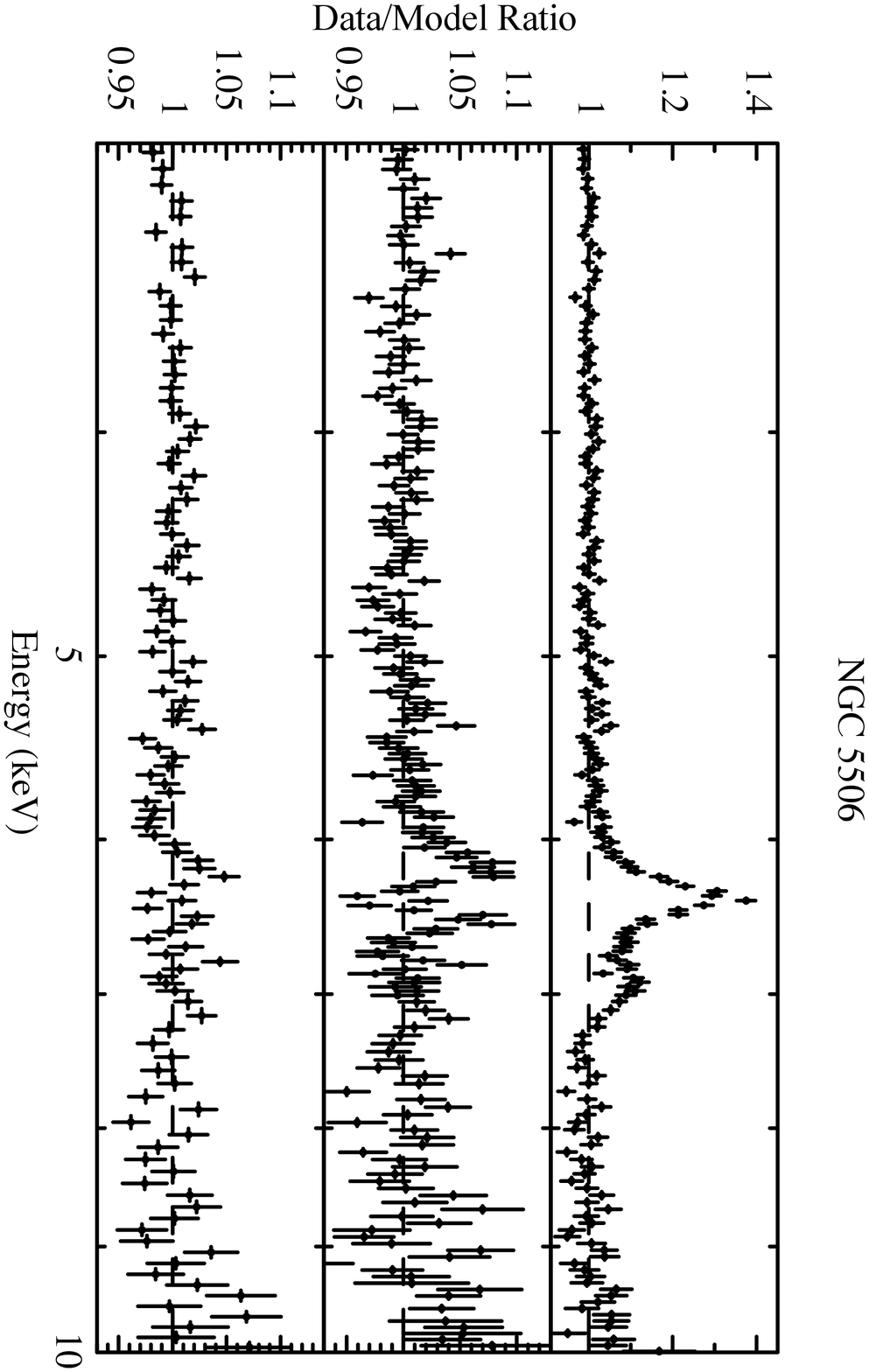}
\end{minipage}
\vspace*{0.15cm}
\hspace*{0.5cm} 
\begin{minipage}[b]{0.4\linewidth}
\centering
\includegraphics[width=0.65\textwidth, angle=90]{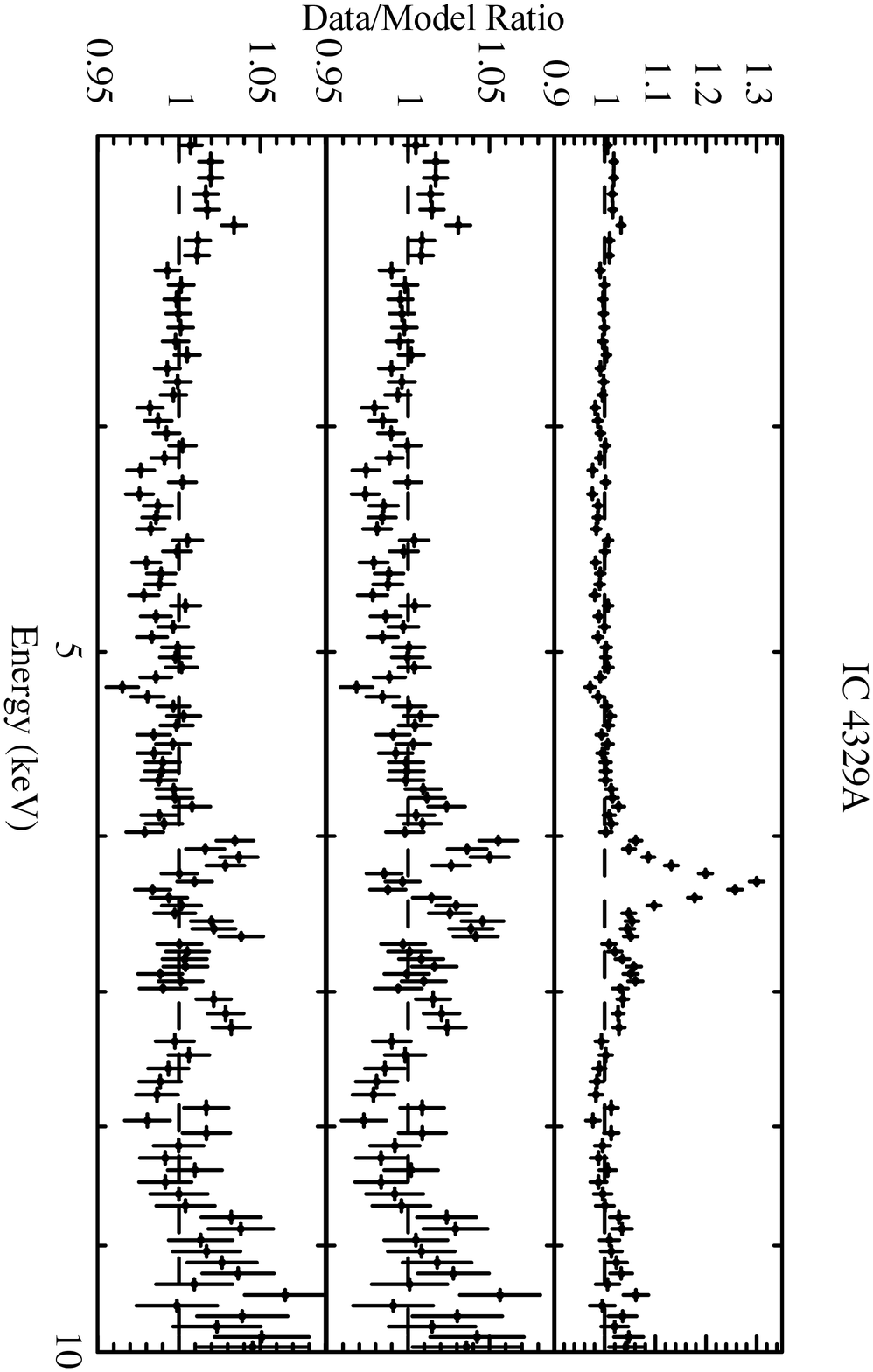}
\end{minipage}
\vspace*{0.15cm}
\hspace*{0.5cm} 
\begin{minipage}[b]{0.4\linewidth}
\centering
\includegraphics[width=0.65\textwidth, angle=90]{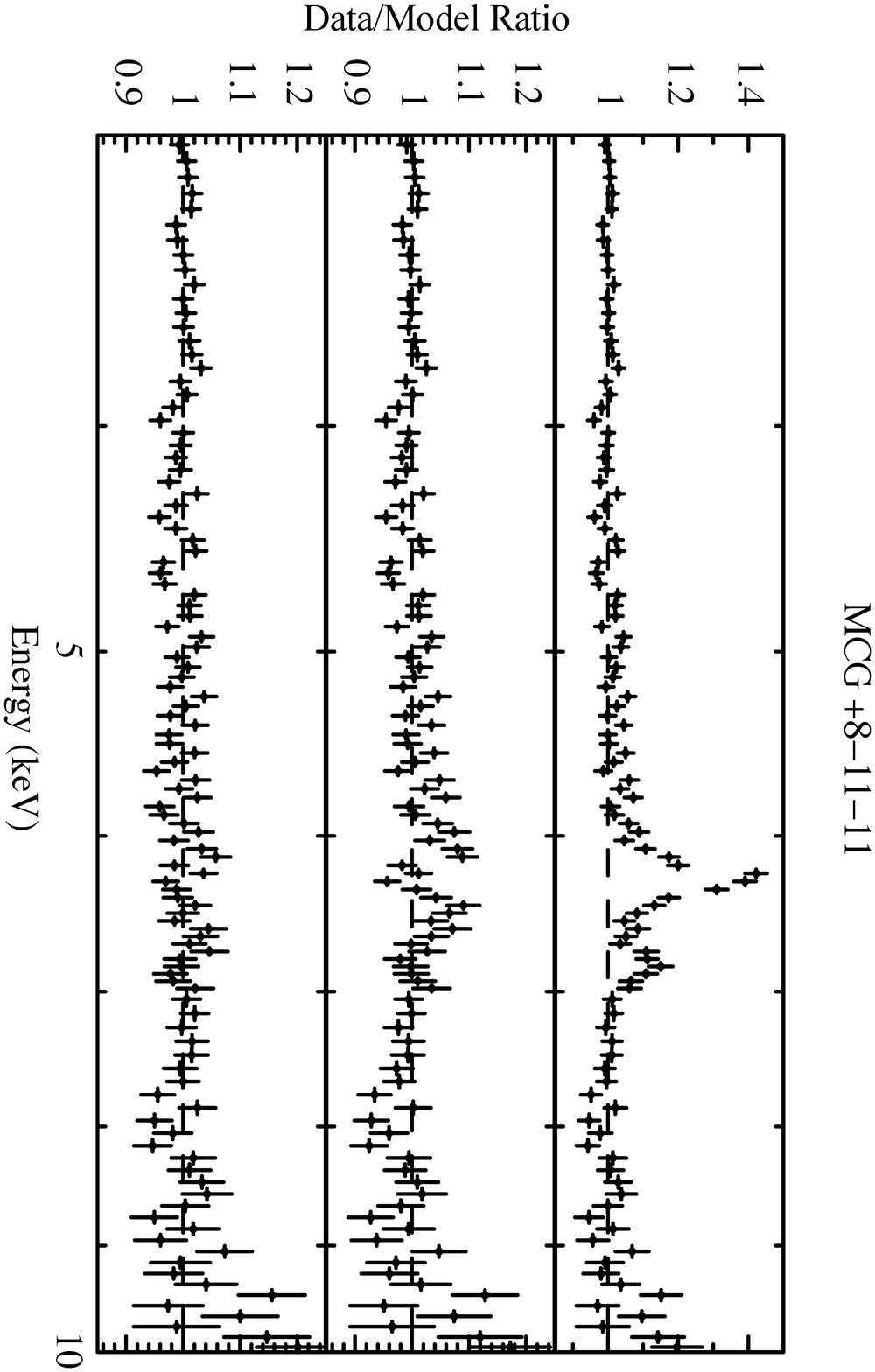}
\end{minipage}
\vspace*{0.15cm}
\hspace*{0.5cm} 
\begin{minipage}[b]{0.4\linewidth}
\centering
\includegraphics[width=0.65\textwidth, angle=90]{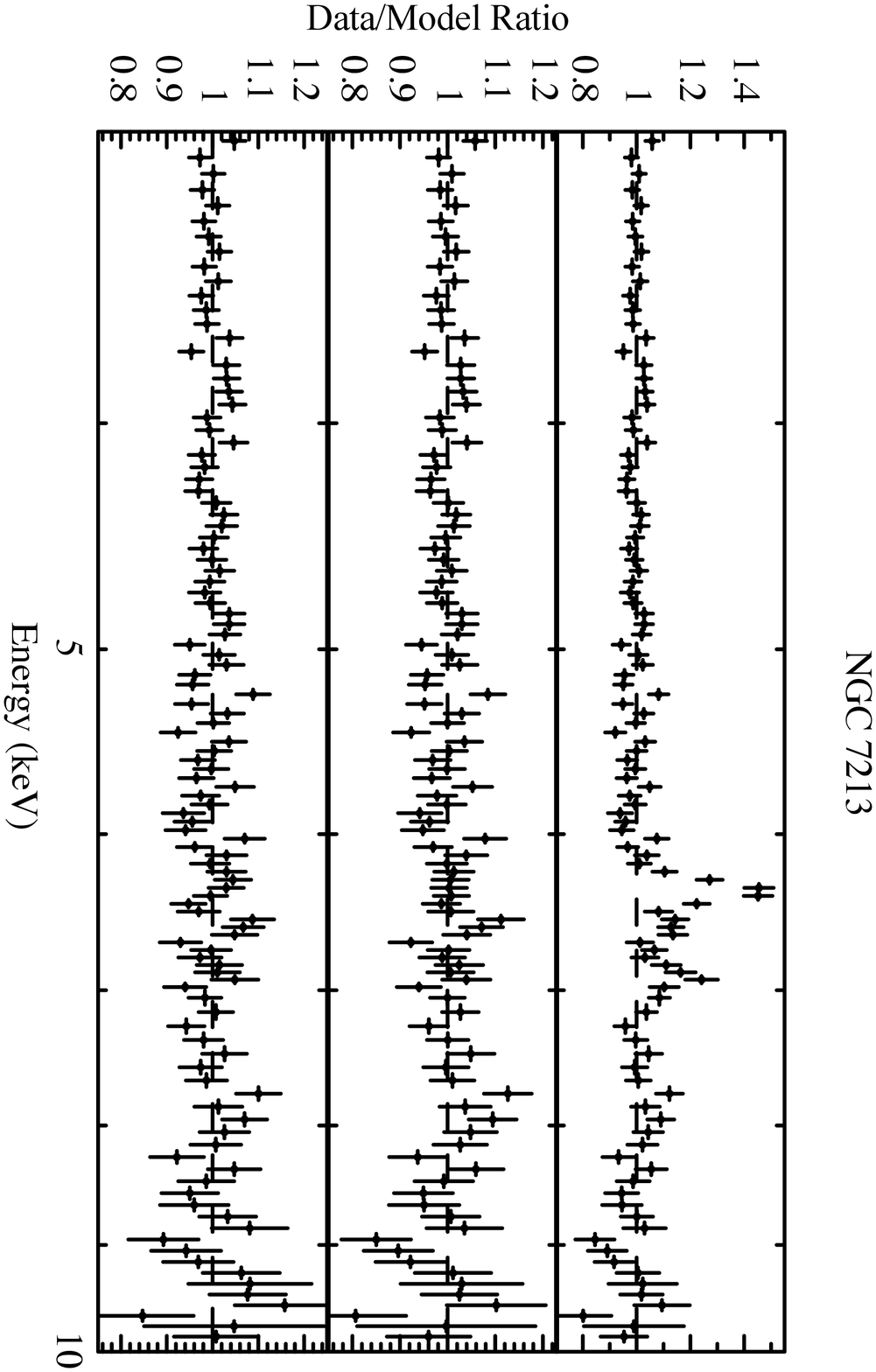}
\end{minipage}
\vspace*{0.15cm}
\hspace*{0.5cm} 
\begin{minipage}[b]{0.4\linewidth}
\centering
\includegraphics[width=0.65\textwidth, angle=90]{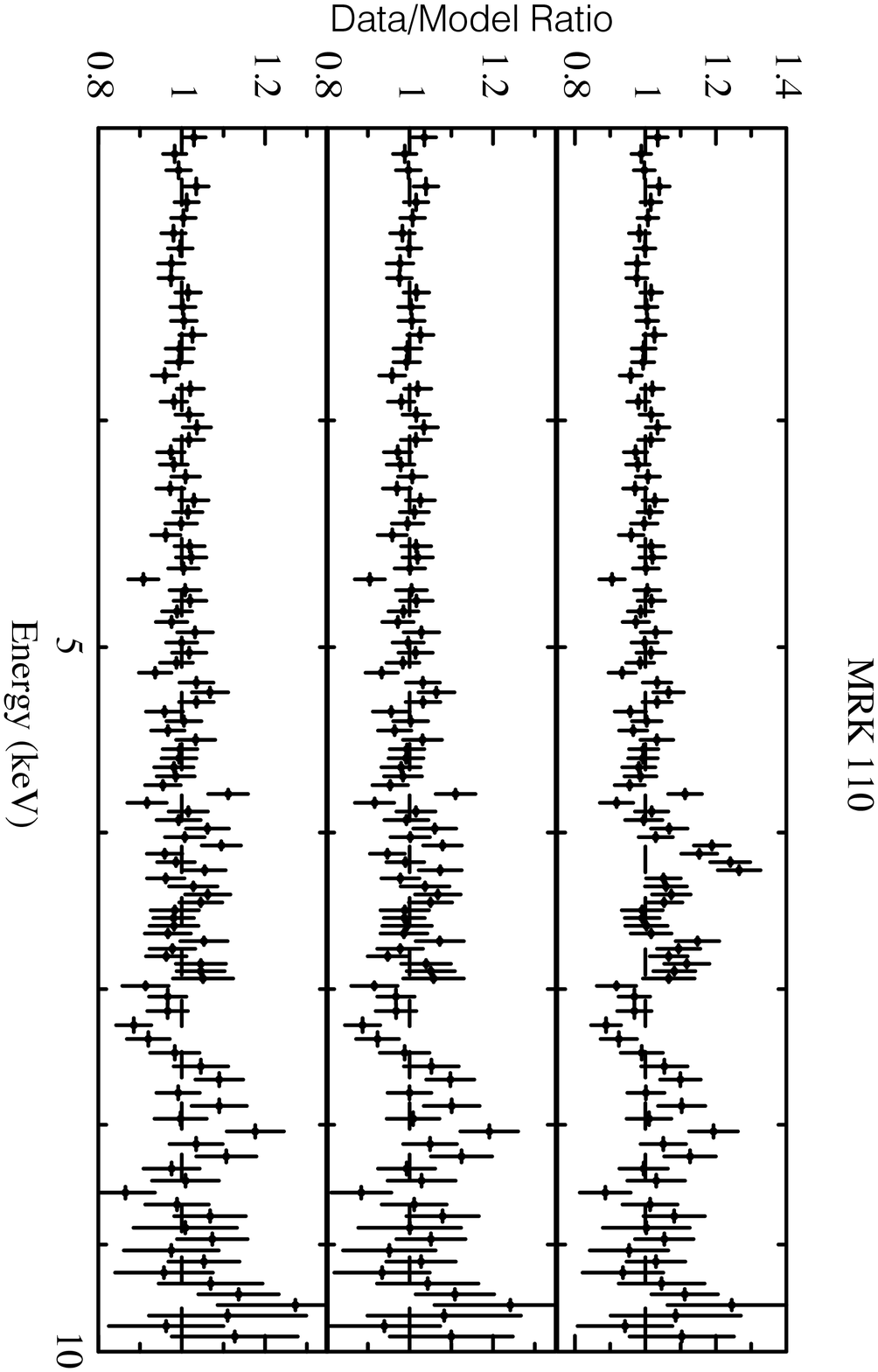}
\end{minipage}
\vspace*{0.15cm}
\hspace*{0.5cm} 
\begin{minipage}[b]{0.4\linewidth}
\centering
\includegraphics[width=0.65\textwidth, angle=90]{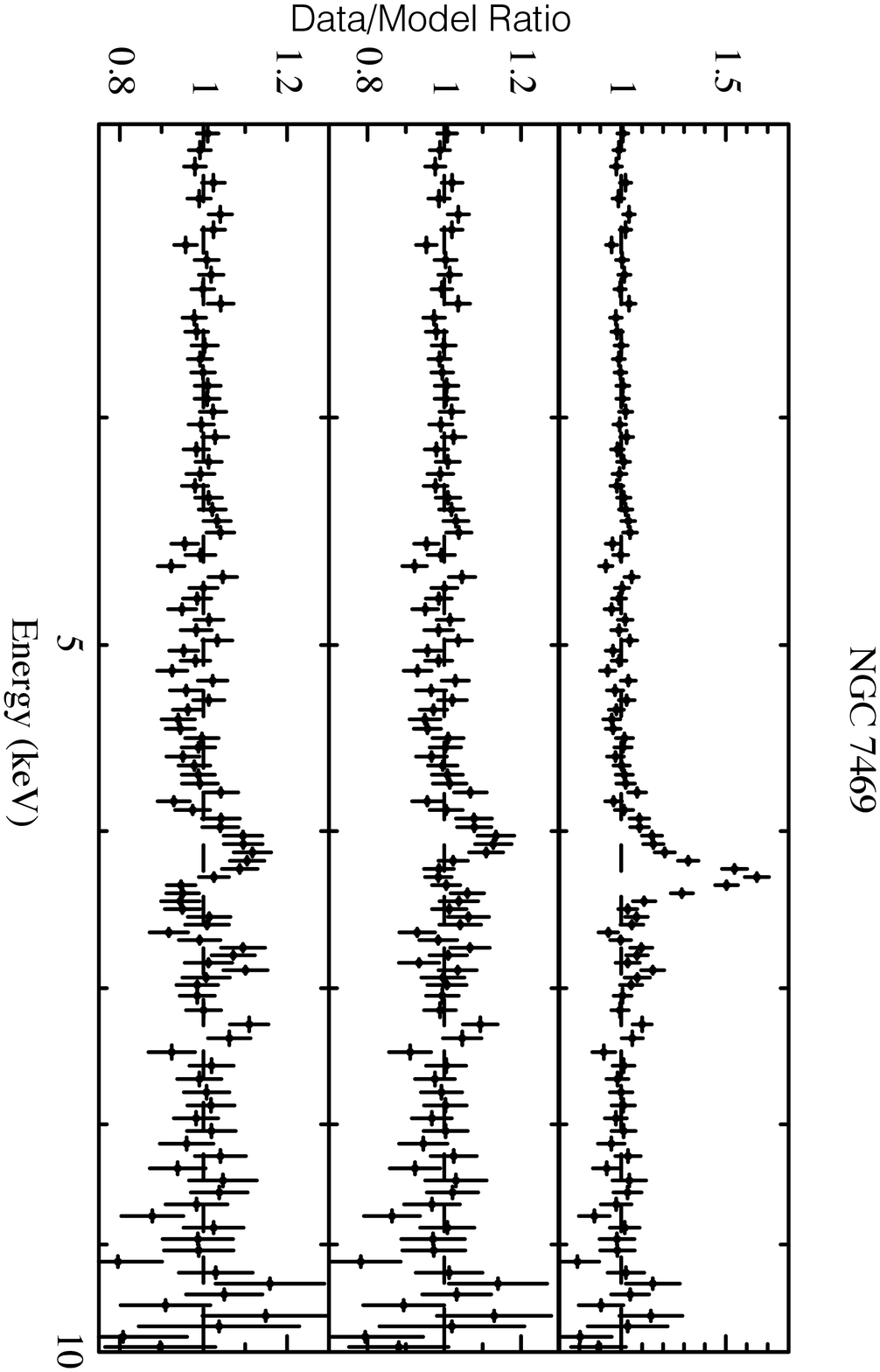}
\end{minipage}
\vspace*{0.15cm}
\hspace*{0.5cm} 
\begin{minipage}[b]{0.4\linewidth}
\centering
\includegraphics[width=0.65\textwidth, angle=90]{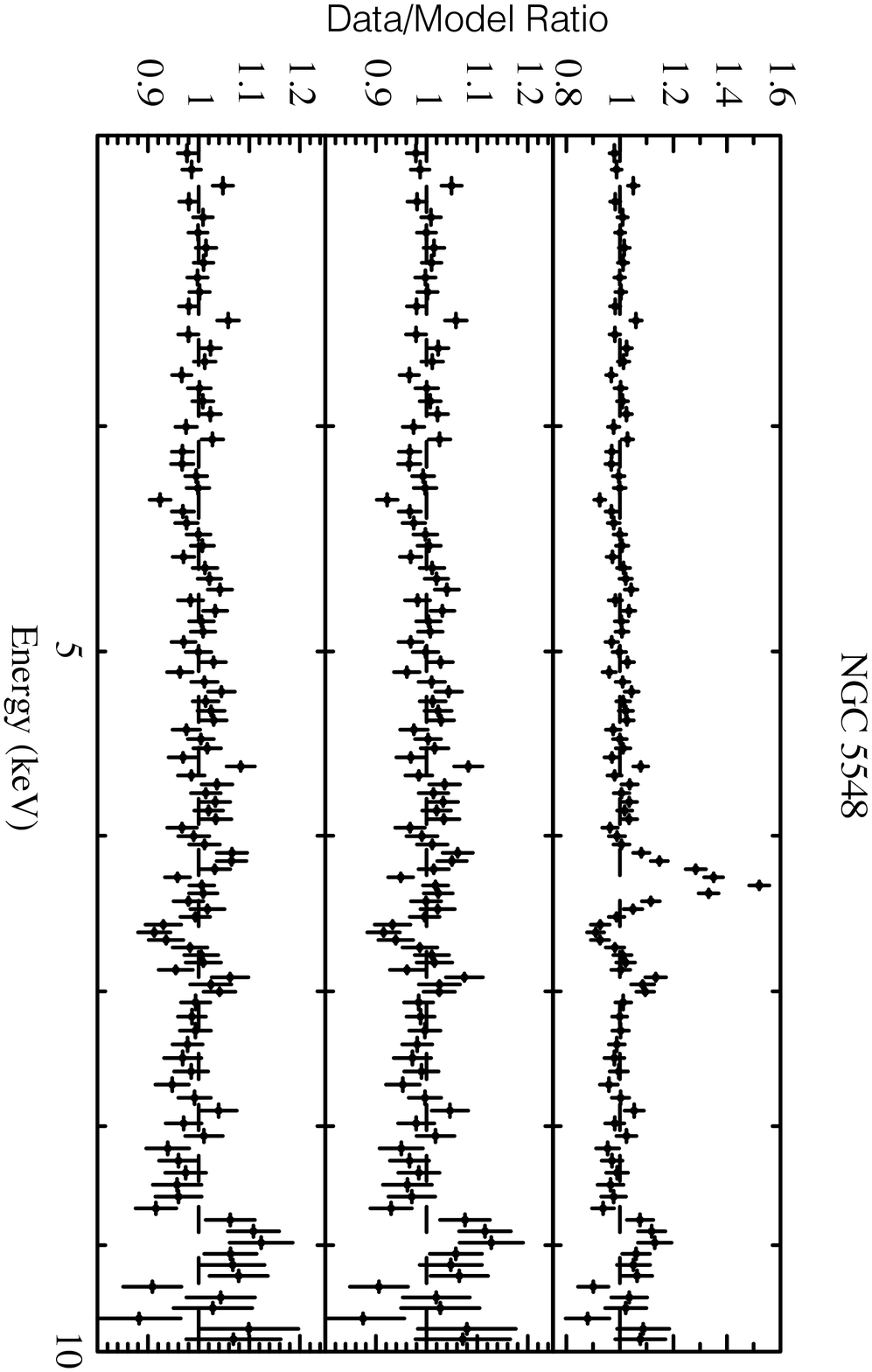}
\end{minipage}
\hspace*{0.5cm} 
\begin{minipage}[b]{0.4\linewidth}
\centering
\includegraphics[width=0.65\textwidth, angle=90]{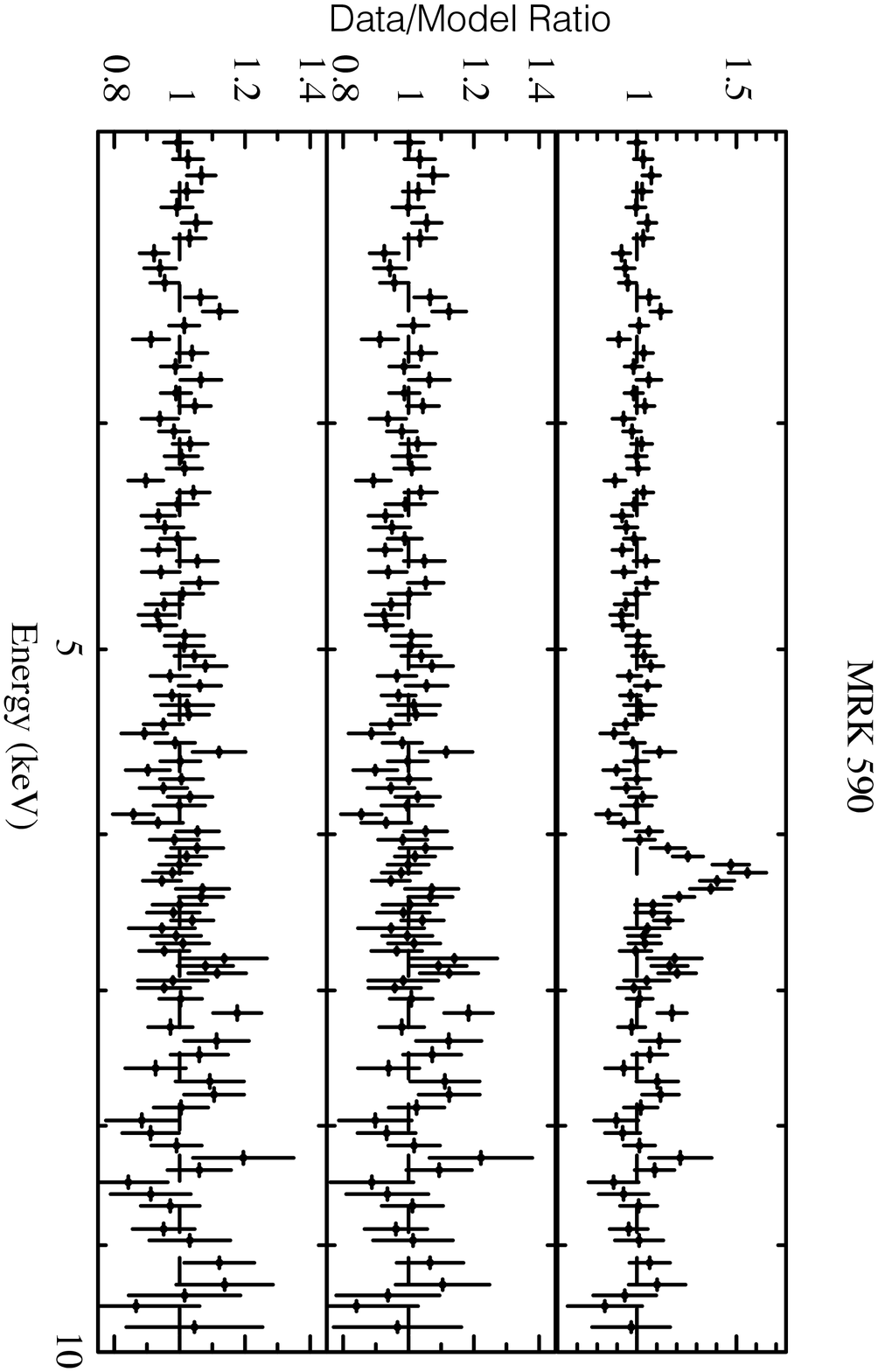}
\end{minipage}
\caption{We present the XIS data to model ratios for each source in the sample. The top panels show a model that includes a neutral absorber at the redshift of the source and a reflection component (as described in Section \ref{Baseline}). The 5-7 keV data have been excluded from these fits. The middle panels show the residuals from a model adding a gaussian component to fit the narrow Fe K$\alpha$ emission line (see Section \ref{lineModel}). The bottom panels show the data to model ratio when a relativistic line line component is included in the previous model (Relativistic Relline Model, see Section \ref{RelLine}). In cases where multiple observations are available for the same object, the data/model ratios have been combined.}
\label{PlotApp}
\end{figure*}

In Figure \ref{PlotApp}, the data to model ratios are presented for each source. The upper panels show the ratios when only the continuum is fitted (which we refer to as the Baseline model). The energy range between 5 and 7 keV was excluded in the fitting process. 

\subsection{Narrow Line Model}
\label{lineModel}

After modelling the primary continuum and the Compton reflection component, there is clear evidence for Fe line emission around 6.4 keV in all sources (see top panels ofFig. \ref{PlotApp}). A narrow Fe component is predicted from any distant material such as a molecular torus, and is commonly observed. To verify this, we added to the Baseline model a narrow gaussian (\textit{zgauss}) in order to fit the main residual at $\sim$6.4 keV (Narrow Line model: \textit{zwabs*(pexrav+zgauss)}), fixing the width to $\sigma$ = 1 eV, while the energy was free to vary. We also included the associated Fe K$\beta$ line. The energy of this was fixed to 7.06 keV, the width to $\sigma$ = 1 eV and the flux was linked to be 11.3 per cent of that of the Fe K$\alpha$ emission line.  

After the inclusion of the narrow Fe K$\beta$ line, we tested for the presence of narrow, ionized emission lines in all the spectra, such as the H-like and He-like lines (Fe~{\sc xxv} and Fe~{\sc xxvi}). For this reason, we added to the previous model two narrow gaussian components. The rest frame energies of these lines were fixed to 6.7 keV and 6.96 keV, respectively, and the widths to $\sigma$ = 1 eV. In Table \ref{LineIron}, we present the results of this test for each observation. Together with the fluxes of the lines, we show the improvements in $\Delta\chi^2$ when those components are included in the model. We conservatively included these lines in the subsequent analysis whenever the 90$\%$ lower limit on the flux was found to be greater than zero. We inferred the presence of both Fe~{\sc xxv} and Fe~{\sc xxvi} lines in all three spectra of NGC 5506, consistent with previous results based on the analysis of \textit{XMM-Newton} observations (\citealt{Matt+01}). In the case of IC 4329A, the Fe~{\sc xxv} emission line was already significantly detected in the previous work of \citet{Mantovani+14}, when the same \textit{Suzaku} observations were analyzed. The residuals from these fits (Narrow Line Model) are shown in the middle panels of Figure \ref{PlotApp}.

\begin{table}
\centering
\caption{Observations where the Fe~{\sc xxv} and the Fe~{\sc xxvi} lines were detected. We also report the improvements in $\Delta\chi^2$/$\Delta$d.o.f. in the cases where the detection of the lines is significant. The fluxes are in 10$^{-5}$ erg s$^{-1}$ cm$^{-2}$.}
\begin{minipage}{150mm}
\label{LineIron}
\resizebox{8.5cm}{!} {
\centering
\begin{tabular}{@{}ccccc@{}}
\hline
Object & Observation & Flux~{\sc xxv} & Flux~{\sc xxvi} & $\Delta\chi^2$/$\Delta$d.o.f.\\
\hline
NGC 5506 & 701030010 & 2.68$_{-0.74}^{+0.73}$ & 2.08$_{-0.74}^{+0.74}$ & 55.5/2\\
& 701030020 & 2.42$_{-0.72}^{+0.72}$ & 1.63$_{-0.72}^{+0.72}$ & 44.68/2\\
& 701030030 & 1.36$_{-0.64}^{+0.64}$ & 1.20$_{-0.64}^{+0.64}$ & 20.62/2\\ \\
IC 4329A & 702113010 & - & 1.44$_{-0.89}^{+0.89}$ & 7.89/1 \\
& 702113020 & - & 1.71$_{-0.88}^{+0.88}$ & 10.12/1\\
& 702113030 & - & - & -\\
& 702113040 & - & - & -\\
& 702113050 & - & 1.10$_{-0.78}^{+0.78}$ & 5.37/1\\
& 707025010 & - & 0.68$_{-0.46}^{+0.46}$ & 6.06/1\\ \\
MCG +8-11-11 & 702112010 & - & 1.22$_{-0.39}^{+0.39}$ & 26.3/1\\ \\
NGC 7213 & 701029010 & 0.69$_{-0.26}^{+0.26}$ & 0.43$_{-0.25}^{+0.25}$ & 22.05/2\\ \\
MRK 110 & 702124010 & - & 0.40$_{-0.24}^{+0.24}$ & 7.72/1\\ \\
NGC 7469 & 703028010 & - & - & -\\ \\
NGC 5548 & 702042010 & - & - & -\\ 
& 702042020 & - & - & - \\
& 702042040 & - & - & -\\
& 702042050 & - & - & -\\
& 702042060 & - & - & -\\
& 702042070 & - & - & -\\
& 702042080 & - & 0.31$_{-0.29}^{+0.30}$ & 3.01/1\\ \\
MRK 590 & 705043010 & 0.29$_{-0.19}^{+0.18}$ &- & 6/1\\
& 705043020 & - & - & -\\ \hline

\end{tabular}}
\end{minipage}
\end{table}

Table \ref{NarrowPar} reports the best fit parameters for each source and observation. The primary Fe K$\alpha$ narrow component is detected in all observations, with equivalent widths ranging from $\sim$40 to $\sim$200 eV. The energy of this component are consistent with the line being produced by neutral material, so the energy of the narrow Gaussian was fixed to 6.4 keV in the subsequent analysis. The slopes of the power law component, reproducing the primary continuum, also vary within the sources of the sample, but the measurements are consistent with values typically observed in Seyfert galaxies (\citealt{Piconcelli+05}, \citealt{Bianchi+09}). 

\begin{table*} 
\centering
\caption{Best-fit parameters for the Narrow Line Model. The column density is in units of 10$^{22}$ cm$^{-2}$ and the normalization of the primary continuum is in units of 10$^{-2}$ photons keV$^{-1}$ cm $^{-2}$ s$^{-1}$. The energy of the Fe line is shown in keV, while the EW is in eV. The FeK$\alpha$ flux is quoted in units of10$^{-5}$ erg s$^{-1}$ cm$^{-2}$. The continuum flux in the 3-10 keV band is in 10$^{-11}$ erg s$^{-1}$ cm$^{-2}$. In the last column we present the best fit $\chi^2$/d.o.f. when the energy of the Fe line is fixed to 6.4 keV.}
\label{NarrowPar}
\resizebox{18cm}{!} {
\begin{tabular}{@{}cccccccccccc@{}}
\hline
Object & Observation & N$_H$ & $\Gamma$ & R & Norm & E$_{K\alpha}$ & Flux$_{K\alpha}$ & EW & Flux$_{3-10 keV}$ & $\chi^2$/d.o.f & $\chi^2$/d.o.f$_{K\alpha}$ \\ 
\hline
NGC 5506 & 701030010 & 2.76$_{-0.22}^{+0.22}$ & 1.92$_{-0.04}^{+0.05}$ & 1.23$_{-0.24}^{+0.27}$ & 4.12$_{-0.27}^{+0.29}$ & 6.41$_{-0.01}^{+0.01}$ & 9.50$_{-0.77}^{+0.78}$ & 72$_{-8}^{+8}$ & 8.59$_{-0.01}^{+0.01}$ & 1362.29/1334 & 1363.73/1335\\
& 701030020 & 3.09$_{-0.20}^{+0.20}$ & 1.95$_{-0.04}^{+0.04}$ & 1.33$_{-0.23}^{+0.26}$ & 4.63$_{-0.28}^{+0.30}$ & 6.39$_{-0.01}^{+0.02}$ & 9.09$_{-0.77}^{+0.77}$ & 64$_{-5}^{+5}$ & 9.12$_{-0.03}^{+0.03}$ & 1463.26/1403 & 1463.89/1404\\
& 701030030 & 3.21$_{-0.19}^{+0.19}$ & 2.02$_{-0.04}^{+0.05}$ & 1.83$_{-0.30}^{+0.35}$ & 4.79$_{-0.30}^{+0.32}$ & 6.39$_{-0.01}^{+0.01}$ & 7.73$_{-0.69}^{+0.68}$ & 58$_{-5}^{+5}$ & 8.55$_{-0.02}^{+0.02}$ & 1411.65/1377 & 1415.12/1378\\ \\
IC 4329A & 702113010 & 0.4 fixed & 1.82$_{-0.03}^{+0.04}$ & 0.79$_{-0.21}^{+0.24}$ & 3.13$_{-0.13}^{+0.15}$ & 6.39$_{-0.02}^{+0.02}$ & 6.26$_{-0.92}^{+0.96}$ & 53$_{-8}^{+8}$ & 8.38$_{-0.01}^{+0.01}$ & 938.11/876 & 939.04/877\\
& 702113020 & 0.4 fixed & 1.92$_{-0.03}^{+0.04}$ & 1.50$_{-0.26}^{+0.30}$ & 4.23$_{-0.17}^{+0.18}$ & 6.39$_{-0.01}^{+0.01}$ & 5.66$_{-0.92}^{+0.92}$ & 40$_{-7}^{+7}$ & 9.85$_{-0.01}^{+0.01}$ & 1091.85/1093& 1092.91/1094\\
& 702113030 & 0.4 fixed & 1.77$_{-0.03}^{+0.03}$ & 0.81$_{-0.19}^{+0.21}$ & 3.27$_{-0.13}^{+0.14}$ & 6.38$_{-0.02}^{+0.01}$ & 7.32$_{-0.99}^{+0.99}$ & 54$_{-7}^{+7}$ & 9.48$_{-0.01}^{+0.01}$ & 985.37/989 & 993.81/990\\
& 702113040 & 0.4 fixed & 1.82$_{-0.04}^{+0.04}$ & 0.97$_{-0.23}^{+0.27}$ & 3.19$_{-0.14}^{+0.15}$ & 6.40$_{-0.02}^{+0.02}$ & 7.58$_{-0.99}^{+1.00}$ & 62$_{-8}^{+8}$ & 8.62$_{-0.01}^{+0.01}$ & 917.53/861& 917.72/862\\
& 702113050 & 0.4 fixed & 1.73$_{-0.05}^{+0.06}$ & 1.51$_{-0.39}^{+0.52}$ & 1.73$_{-0.11}^{+0.13}$ & 6.40$_{-0.02}^{+0.02}$ & 5.65$_{-0.82}^{+0.84}$  & 69$_{-10}^{+10}$ & 5.65$_{-0.02}^{+0.02}$ & 589.74/586 & 589.83/587\\
& 707025010 & 0.4 fixed & 1.74$_{-0.01}^{+0.01}$ & 0.48$_{-0.07}^{+0.08}$ & 2.88$_{-0.05}^{+0.05}$ & 6.40$_{-0.01}^{+0.01}$ & 7.88$_{-0.48}^{+0.48}$ & 65$_{-4}^{+4}$ & 8.65$_{-0.01}^{+0.01}$ & 1688.70/1639 & 1689.37/1640\\ \\
MCG +8-11-11 & 702112010 & - & 1.65$_{-0.01}^{+0.02}$ & $<$0.12 & 1.49$_{-0.03}^{+0.04}$ & 6.39$_{-0.01}^{+0.01}$ & 5.61$_{-0.42}^{+0.42}$ & 78$_{-6}^{+6}$ & 5.19$_{-0.01}^{+0.01}$ & 1296.96/1324 & 1298.25/1325\\ \\
NGC 7213 & 701029010 & - & 1.75$_{-0.07}^{+0.07}$ & $<$ 1.21 & 0.62$_{-0.04}^{+0.05}$ & 6.39$_{-0.01}^{+0.01}$ & 2.08$_{-0.25}^{+0.28}$ & 81$_{-11}^{+11}$ & 1.89$_{-0.01}^{+0.01}$ & 677.09/689 & 680.17/690 \\ \\
MRK 110 & 702124010 & - & 1.74$_{-0.04}^{+0.04}$ & $<$ 0.41 & 0.55$_{-0.03}^{+0.03}$ & 6.39$_{-0.03}^{+0.03}$ & 1.06$_{-0.24}^{+0.25}$ & 44$_{-10}^{+10}$ & 1.65$_{-0.01}^{+0.01}$ & 546.96/546 & 547.12/547\\ \\
NGC 7469 & 703028010 & - & 1.75$_{-0.05}^{+0.05}$ & 0.94$_{-0.32}^{+0.40}$ & 0.52$_{-0.03}^{+0.03}$ & 6.38$_{-0.01}^{+0.01}$ & 2.73$_{-0.24}^{+0.24}$ & 119$^{+10}_{-11}$ & 1.64$_{-0.01}^{+0.01}$ & 633.72/667 & 650.95/668\\ \\
NGC 5548 & 702042010 & - & 1.45$_{-0.11}^{+0.22}$ & $<$3.58 & 0.12$_{-0.02}^{+0.03}$ & 6.40$_{-0.02}^{+0.02}$ & 1.85$_{-0.33}^{+0.33}$ & 192$^{+34}_{-34}$ & 0.68$_{-0.01}^{+0.01}$ & 83.72/101 & 83.87/102\\ 
& 702042020 & - & 1.68$_{-0.13}^{+0.14}$ & $<$2.78 & 0.30$_{-0.04}^{+0.05}$ & 6.39$_{0.02}^{+0.02}$ & 2.03$_{-0.34}^{+0.35}$ & 132$_{-23}^{+23}$ & 1.10$_{-0.01}^{+0.01}$ & 235.1/188 & 236.91/189\\
& 702042040 & - & 1.69$_{-0.06}^{+0.07}$ & $<$0.71 & 0.61$_{-0.05}^{+0.05}$ & 6.37$_{-0.03}^{+0.03}$ & 1.82$_{-0.43}^{+0.43}$ & 64$^{+15}_{-15}$ & 2.04$_{-0.02}^{+0.02}$ & 326.84/312 & 328.70/313\\
& 702042050 & - & 1.55$_{-0.08}^{+0.09}$ & 0.79$_{-0.49}^{+0.68}$ & 0.33$_{-0.03}^{+0.04}$ & 6.38$_{-0.02}^{+0.02}$ & 1.90$_{-0.39}^{+0.39}$ & 92$_{-19}^{+19}$ & 1.45$_{-0.01}^{+0.01}$ & 220.66/230 & 222.41/231\\
& 702042060 & - & 1.61$_{-0.05}^{+0.05}$ & $<$0.49 & 0.70$_{-0.04}^{+0.05}$ & 6.41$_{-0.03}^{+0.03}$ & 1.80$_{-0.49}^{+0.49}$ & 49$_{-13}^{+13}$ & 2.77$_{-0.03}^{+0.03}$ & 388.77/374 & 389.54/375\\
& 702042070 & - & 1.58$_{-0.06}^{+0.07}$ & 0.42$_{-0.32}^{+0.42}$ & 0.45$_{-0.03}^{+0.04}$ & 6.39$_{-0.02}^{+0.02}$ & 2.28$_{-0.42}^{+0.42}$ & 90$_{-17}^{+17}$& 1.80$_{-0.01}^{+0.01}$ & 252.40/295 & 253.00/296\\
& 702042080 & - & 1.63$_{-0.15}^{+0.16}$ & $<$3.38 & 0.24$_{-0.04}^{+0.04}$ & 6.41$_{-0.02}^{+0.01}$ & 2.16$_{-0.32}^{+0.36}$ & 158$_{-26}^{+26}$ &0.98$_{-0.01}^{+0.01}$ & 170.68/180 & 172.86/181\\ \\
MRK 590 & 705043010 & - & 1.65$_{-0.05}^{+0.14}$ & $<$1.39 & 0.17$_{-0.01}^{+0.03}$ & 6.40$_{-0.02}^{+0.02}$ & 1.06$_{-0.21}^{+0.20}$ & 118$_{-25}^{+24}$ & 0.62$_{-0.01}^{+0.01}$ & 199.99/186 & 200.44/187\\
& 705043020 & - & 1.56$_{-0.11}^{+0.20}$ & $<$2.74 & 0.13$_{-0.02}^{+0.03}$ & 6.41$_{-0.03}^{+0.03}$ & 0.86$_{-0.25}^{+0.25}$ & 106$_{-31}^{+31}$ & 0.57$_{-0.01}^{+0.01}$ & 103.76/116& 104.35/117\\ \hline
\end{tabular}}
\end{table*}

\subsection{Relativistic FeK$\alpha$ analysis}
\label{RelLine}

We next investigated if a relativistic Fe K$\alpha$ line improves the quality of the fits presented in Table \ref{NarrowPar}. Ideally one would compare the $\chi^2$ of the fit with and without the broad line component and estimate its significance. However, assessing the significance of relativistic emission lines in the X-ray spectra of AGN is not straightforward. For example, it has been pointed out that the F-test is not a precise tool to estimate the chance probability of detecting these or similar features (\citealt{Protassov+02}). 

\subsubsection{Estimating the significance of broad Fe line detections for a given $\Delta\chi^2$}
In order to estimate the appropriate significance of the line given a $\Delta\chi^2$ value, we performed Monte Carlo simulations. We started by simulating six thousand spectra. The model used in this process includes a cut-off power law, a reflection component with narrow Fe K$\alpha$, Fe K$\beta$ and Ni K$\alpha$ emission lines and an Fe Compton shoulder: \textit{cutoffpl+pexmon}. We used the \textit{xspec} \textit{fakeit} command with the addition of the appropriate noise. We took into account the response files of the XIS instrument (i.e. XIS$\_$FI.rmf and XRT$\_$FI$\_$xisnom.arf files). Once the spectra were simulated, we first refitted them with the \textit{cutoffpl+pexmon} model used to simulate them, and then we added a \textit{relline} component (\citealt{Dauser+10}). For each simulated spectrum, we recorded the best fits $\chi^2$ for the two different models. By recording how many times a $\Delta\chi^2$ better than the observed values happened by chance, we can establish a relation between the observed $\Delta\chi^2$ and the significance of the detection of the\textit{relline} component in the real data. 

We repeated this process using different relativistic line profiles in order to test if the threshold value of $\Delta\chi^2$depends on the parameters adopted in the \textit{relline} model. Since it is difficult to constrain the most important parameters of the relativistic line, especially in the cases where the line is weak and/or very broad, we fixed the following parameters of the relline model: i) E$_{Relline}$ = 6.4 keV; ii) index1 = index2 = 3; iii) R$_{break}$ = 15 r$_g$; iv) R$_{out}$ = 400 r$_g$; v) limb = 0. Since the sources of the sample are nearby objects, we assumed $z=0$. This allows us to obtain simulations that are useful for sources at low redshift. To correspond to the fits to the actual data (see below), we fitted the line profile using different fixed inclinations (30$^{\circ}$, 60$^{\circ}$, 80$^{\circ}$) and inner radii (6 r$_g$ and 1.24 r$_g$). When the inner radius was 6 r$_g$ we used a spin parameter a=0, appropriate for a Schwarzschild black hole. When it was fixed to 1.24 r$_g$, the spin parameter was changed to a = 0.998 corresponding to a maximally spinning Kerr black hole. Table \ref{chi} presents the results of the simulations. The $\Delta\chi^2$ values reported correspond to the 95$\%$ confidence level for each combination of inclination and inner radius.

\begin{table}
\caption{Results of the simulations for $\Delta\chi^2$ at the 95$\%$ confidence level.The values are presented for each combination of inclination and inner radius.}
\begin{minipage}{85mm}
\label{chi}
\begin{center}
\begin{tabular}{@{}ccc@{}}
\hline
$\theta$ & R$_{in}$ & $\Delta\chi^2$(95$\%$)\\
\hline
30$^{\circ}$ & 6 r$_g$ & 4.71\\
30$^{\circ}$ & 1.24 r$_g$ & 5.57\\
60$^{\circ}$ & 6 r$_g$ & 2.67\\
60$^{\circ}$ & 1.24 r$_g$ & 2.99\\
80$^{\circ}$ & 6 r$_g$ & 1.99\\
80$^{\circ}$ & 1.24 r$_g$ & 1.44\\ \hline
\end{tabular}
\end{center}
\end{minipage}
\end{table}

\subsubsection{Spectral Analysis}

\begin{table*}
\centering
\caption{Best-fit parameters for the Relativistic Relline model. The column density is in units of 10$^{22}$ cm$^{-2}$ and the normalization of the primary continuum is in units of 10$^{-2}$ photons keV$^{-1}$ cm $^{-2}$ s$^{-1}$. The Fe K$\alpha$ fluxes are in units of 10$^{-5}$ erg s$^{-1}$ cm$^{-2}$, and the continuum flux in the 3-10 keV band is in units of 10$^{-11}$ erg s$^{-1}$ cm$^{-2}$.}
\label{relline}
\resizebox{18cm}{!} {
\begin{tabular}{@{}ccccccccccccccc@{}}
\hline
Object & Observation & N$_H$ & $\Gamma$ & R & Norm & Flux$_{K\alpha}$ & EW & Flux$_{3-10 keV}$ & $\theta_{disk}$ & R$_{in}$ & EW$_{Rell}$& $\chi^2$/d.o.f & $\Delta\chi^2/\Delta d.o.f.$& Significance\\
\hline
NGC 5506 & 701030010 & 2.35$_{-0.17}^{+0.17}$ & 1.87$_{-0.03}^{+0.03}$ & 1.00$_{-0.14}^{+0.15}$ & 3.71$_{-0.17}^{+0.18}$ & 8.26$_{-0.57}^{+0.57}$ & 61$_{-4}^{+4}$ & 8.60$_{-0.02}^{+0.02}$ & 30$^{\circ}$ & 6r$_g$ & 62$_{-16}^{+16}$& 1348.65/1334 & 15.08/1 & 99.998$\%$ \\
& 701030020 & 2.50$_{-0.15}^{+0.15}$ & 1.87$_{-0.03}^{+0.03}$ & 0.99$_{-0.12}^{+0.13}$ & 3.97$_{-0.17}^{+0.18}$ & 7.16$_{-0.55}^{+0.55}$ & 49$_{-4}^{+4}$ & 9.13$_{-0.02}^{+0.02}$ & 30$^{\circ}$ & 6r$_g$ & 96$_{-15}^{+15}$ & 1423.86/1403 & 40.03/1& $>$ 99.999 $\%$  \\
& 701030030 &2.64$_{-0.15}^{+0.15}$ & 1.93$_{-0.03}^{+0.03}$ & 1.35$_{-0.16}^{+0.17}$ & 4.09$_{-0.18}^{+0.19}$ & 6.34$_{-0.47}^{+0.48}$ & 46$_{-3}^{+3}$ & 8.54$_{-0.02}^{+0.02}$ & 30$^{\circ}$ & 6r$_g$ & 89$_{-15}^{+15}$ & 1381.03/1377 & 34.09/1& $>$ 99.99 $\%$ \\ \\
IC 4329A & 702113010 & 0.4 fixed & 1.83$_{-0.02}^{+0.02}$ & 0.84$_{-0.13}^{+0.15}$ & 3.16$_{-0.08}^{+0.09}$ & 5.07$_{-0.65}^{+0.65}$ & 41$_{-5}^{+5}$& 8.37$_{-0.01}^{+0.01}$ & 30$^{\circ}$ & 6r$_g$ & 69$_{-18}^{+18}$ & 923.58/876 & 15.46/1 & 99.986 $\%$  \\
& 702113020 & 0.4 fixed & 1.92$_{-0.01}^{+0.02}$ & 1.50$_{-0.16}^{+0.17}$ & 4.23$_{-0.09}^{+0.09}$ & 5.63$_{-0.56}^{+0.55}$ & 40$_{-4}^{+4}$ & 9.85$_{-0.02}^{+0.02}$ & 30$^{\circ}$ & 6r$_g$ & $<$24 & 1092.91/1093 &-  &-\\
& 702113030 & 0.4 fixed & 1.78$_{-0.02}^{+0.02}$ & 0.83$_{-0.12}^{+0.13}$ & 3.29$_{-0.08}^{+0.08}$ & 6.49$_{-0.67}^{+0.67}$ & 47$_{-5}^{+5}$ & 9.48$_{-0.01}^{+0.01}$ & 30$^{\circ}$ & 6r$_g$ & $<$ 66 & 990.03/989 & - & -\\
& 702113040 & 0.4 fixed & 1.83$_{-0.02}^{+0.02}$ & 1.01$_{-0.14}^{+0.16}$ & 3.21$_{-0.09}^{+0.09}$ & 6.62$_{-0.68}^{+0.68}$ & 53$_{-5}^{+5}$ & 8.61$_{-0.01}^{+0.01}$ & 30$^{\circ}$ & 6r$_g$ & 54$_{-18}^{+18}$ & 908.35/861 & 9.37/1 & 99.704 $\%$ \\
& 702113050 & 0.4 fixed & 1.74$_{-0.03}^{+0.03}$ & 1.52$_{-0.24}^{+0.28}$ & 1.73$_{-0.07}^{+0.07}$ & 5.08$_{-0.57}^{+0.57}$ & 61$_{-7}^{+7}$ & 5.65$_{-0.01}^{+0.01}$ & 30$^{\circ}$ & 6r$_g$ & $<$ 90 & 585.23/586 & - & -\\
& 707025010 & 0.4 fixed & 1.75$_{-0.09}^{+0.09}$ & 0.53$_{-0.05}^{+0.05}$ & 2.91$_{-0.03}^{+0.03}$ & 6.80$_{-0.35}^{+0.35}$ & 54$_{-3}^{+3}$ & 8.64$_{-0.02}^{+0.02}$ & 30$^{\circ}$ & 6r$_g$ & 49$_{-8}^{+8}$ & 1656.27/1639 & 33.1/1 & $>$ 99.999 $\%$ \\ \\
MCG +8-11-11 & 702112010 & - & 1.68$_{-0.01}^{+0.01}$ & 0.12$_{-0.05}^{+0.06}$ & 1.52$_{-0.02}^{+0.03}$ & 4.59$_{-0.29}^{+0.29}$ & 60$_{-4}^{+4}$ & 5.17$_{-0.01}^{+0.01}$ & 30$^{\circ}$ & 6r$_g$ & 96$_{-12}^{+12}$ & 1236.05/1324 & 62.2/1 & $>$ 99.999 $\%$ \\ \\
NGC 7213 & 701029010 & -& 1.85$_{-0.06}^{+0.06}$ & 0.82$_{-0.55}^{+0.60}$ & 0.70$_{-0.04}^{+0.05}$ & 1.98$_{-0.17}^{+0.17}$ & 77$_{-6}^{+6}$ & 1.89$_{-0.01}^{+0.01}$ & 80$^{\circ}$ & 6r$_g$ & 233$_{-80}^{+80}$ & 671.73/689 & 8.44/1 & $>$ 99.999 $\%$  \\ \\
MRK 110 & 702124010 & - & 1.79$_{-0.04}^{+0.05}$ & 0.32$_{-0.22}^{+0.26}$ & 0.58$_{-0.03}^{+0.03}$ & 1.05$_{-0.15}^{+0.15}$ & 44$_{-66}^{+6}$ & 1.65$_{-0.01}^{+0.01}$ & 80$^{\circ}$ & 1.24r$_g$ & 148$_{-122}^{+122}$ & 545.66/546 & 1.46/1 & 95.110$\%$ \\ \\
NGC 7469 & 703028010 & - & 1.76$_{-0.03}^{+0.03}$ & 0.96$_{-0.20}^{+0.22}$ & 0.53$_{-0.02}^{+0.02}$ & 2.39$_{-0.16}^{+0.16}$ & 100$_{-7}^{+7}$ & 1.64$_{-0.01}^{+0.01}$ & 30$^{\circ}$ & 6r$_g$ & 73$_{-20}^{+20}$ & 637.81/667 & 13.14/1 & 99.977 $\%$ \\ \\
NGC 5548 & 702042010 & - & 1.55$_{-0.16}^{+0.17}$ & 1.56$_{-1.34}^{+1.70}$ & 0.14$_{-0.02}^{+0.03}$ & 1.84$_{-0.20}^{+0.20}$ & 190$_{-21}^{+21}$ & 0.68$_{-0.01}^{+0.01}$ & 80$^{\circ}$ & 1.24r$_g$ & $<$680 & 83.01/101 & - & - \\ 
& 702042020 & - & 1.82$_{-0.12}^{+0.12}$ & 1.79$_{-1.00}^{+1.20}$ & 0.35$_{-0.04}^{+0.05}$ & 2.00$_{-0.21}^{+0.21}$ & 130$_{-14}^{+14}$ & 1.09$_{-0.01}^{+0.01}$ & 80$^{\circ}$ & 1.24r$_g$ & 489$_{-249}^{+247}$ & 233.10/188 & 3.81/1 & 99.350 $\%$   \\
& 702042040 & - & 1.69$_{-0.03}^{+0.05}$ & 0.26$_{-0.24}^{+0.28}$ & 0.61$_{-0.03}^{0.04}$ & 1.79$_{-0.26}^{+0.26}$ & 63$_{-9}^{+9}$ & 2.04$_{-0.01}^{+0.01}$ & 80$^{\circ}$ & 1.24r$_g$ & $<$131 & 328.70/312 & - & -\\
& 702042050 & - & 1.58$_{-0.07}^{+0.10}$ & 0.96$_{-0.43}^{+0.66}$ & 0.34$_{-0.03}^{+0.04}$ & 1.86$_{-0.24}^{+0.24}$ & 90$_{-12}^{+12}$ & 1.45$_{-0.01}^{+0.01}$ & 80$^{\circ}$ & 1.24r$_g$ & $<$317 & 222.19/230 &- & -\\
& 702042060 & - & 1.61$_{-0.04}^{+0.02}$ & 0.20$_{-0.13}^{+0.17}$ & 0.70$_{-0.03}^{+0.02}$ & 1.77$_{-0.30}^{+0.30}$ & 48$_{-8}^{+8}$ & 2.65$_{-0.01}^{+0.01}$ & 80$^{\circ}$ & 1.24r$_g$ & $<$43 & 389.54/374 &- & -\\
& 702042070 & - & 1.58$_{-0.04}^{+0.04}$ & 0.42$_{-0.24}^{+0.24}$ & 0.45$_{-0.02}^{+0.02}$ & 2.27$_{-0.25}^{+0.25}$ & 90$_{-10}^{+10}$ & 1.80$_{-0.01}^{0.01}$ & 80$^{\circ}$ & 1.24r$_g$ & $<$105 & 253.00/295 &- &\\
& 702042080 & - & 1.63$_{-0.10}^{+0.10}$ & 1.43$_{-0.60}^{+0.60}$ & 0.24$_{-0.03}^{+0.02}$ & 2.15$_{-0.19}^{+0.19}$ & 156$_{-14}^{+14}$ & 0.98$_{-0.02}^{+0.02}$ & 80$^{\circ}$ & 1.24r$_g$ & $<$113 & 172.86/180 &- & -\\ \\
MRK 590 & 705043010 & - & 1.68$_{-0.05}^{+0.12}$ & $<$ 1.04 & 0.18$_{-0.01}^{+0.03}$ & 1.06$_{-0.13}^{+0.12}$ & 118$_{-15}^{+13}$ & 0.61$_{-0.03}^{0.03}$ & 80$^{\circ}$ & 1.24r$_g$ & $<$ 494 & 198.71/186 &- & -\\
& 705043020 & - & 1.70$_{-0.15}^{+0.17}$ & $<$2.70 & 0.17$_{-0.02}^{+0.03}$ & 0.83$_{-0.15}^{+0.15}$ & 102$_{-19}^{+19}$ & 0.57$_{-0.02}^{+0.02}$ & 80$^{\circ}$ & 1.24r$_g$ & 465$_{-329}^{+324}$ & 102.35/116 & 2/1 & 96.750$\%$ \\ \hline 
\end{tabular}}
\end{table*}

Having determined the critical $\Delta\chi^2$ values, we then fitted all the actual source spectra using the Relativistic Relline model (RR): \textit{zwabs*(pexrav+zgauss+relline)} with the same \textit{relline} parameters listed above, and the same combinations of inner radius and inclination. These fixed combinations are necessary because it is very difficult to constrain all parameters of the relativistic line simultaneously, bearing in mind that the spectra in our sample have been chosen specifically because the relativistic component is weak and/or of low significance. The Relativistic Relline model includes a neutral absorber at the redshift of the source (\textit{zwabs}), a cutoff power law together with a reflection continuum (\textit{pexrav}), a narrow emission line (\textit{zgauss}) and a relativistic line component (\textit{relline}). The width of the narrow gaussian was fixed to $\sigma$ = 1 eV, and the energy to 6.4 keV. We also included the corresponding narrow Fe K$\beta$. Fe~{\sc xxv} and Fe~{\sc xxvi} emission lines were included when significantly detected in the prior analysis (see Table \ref{LineIron}). For each source, we fitted all the combinations of inclinations and inner radii. Thereafter, we selected all the solutions which lead to an improvement in $\chi^2$ at more than 95$\%$ (see Table \ref{chi} for the $\Delta\chi^2$ values adopted). If more than one solution satisfies this criterion, we choose the one with the highest $\Delta\chi^2$. In cases where the line is not detected, we have chosen the combination which leads to the best fit with the highest $\Delta\chi^2$ in order to calculate the upper limits for the relativistic Fe line fluxes. The best fit parameters for the Relativistic Relline model are presented in Table \ref{relline}. The errors correspond to 1$\sigma$ confidence level, whereas upper limits correspond to the 95$\%$ confidence level as determined from the simulations. The relativistic Fe K$\alpha$ emission line is detected at least in one observation of every source in our sample, and in a total of 12 out of 22 observations. In cases where the significance exceeds 95$\%$ confidence, we determine the significance more precisely using the distributions of $\Delta\chi^2$ from the simulations described above. We report these values in the last column of Table \ref{relline}. We also present the improvements in $\Delta\chi^2/d.o.f.$ when the \textit{relline} model is included in the Narrow Line model. The mean value for the equivalent width of this component averaged over all the sample is $\sim$ 100 eV. 

\begin{figure}
\begin{minipage}[h!]{0.9\linewidth}
\centering
\includegraphics[width=0.7\textwidth, angle=90]{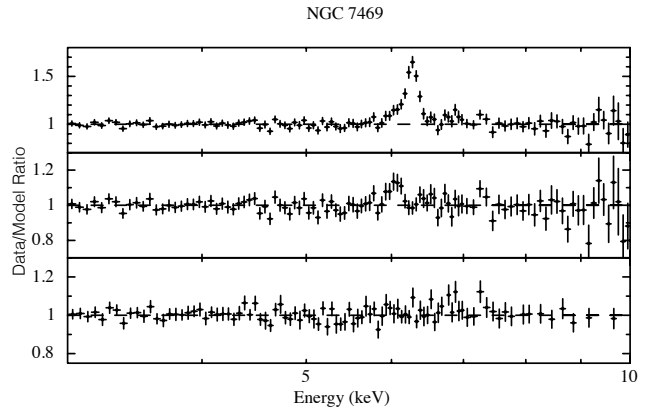}
\end{minipage}
\caption{Data to model ratio of the XIS spectrum for NGC 7469. The Baseline model (upper panel), the Narrow Line model (middle panel) and the Relativistic Relline model (bottom panel) are applied to the data. In the Relativistic Relline model the inclination, the inner radius and the emissivity index are now free to vary in the fit.}
\label{7469}
\end{figure}

The bottom panels of Figure \ref{PlotApp} show the data to model ratios for the XIS spectra when the Relativistic Relline model is applied to the data. The broad line residuals present in several cases in the middle panels (Narrow Line model) are generally well fitted by the \textit{relline} model component. However, in the case of NGC 7469 residuals around 6.4 keV are still present even if a relativistic Fe line is included (and detected at more than 95$\%$ confidence). This could be due to the fact that we have used fixed parameters in the fits. We tested this by letting the inner radius, the inclination and the emissivity index free to vary in the \textit{relline} model component for this source. We found a much better fit in this case with a $\chi^2/d.o.f.$ = 601.13/665, compared to that with frozen parameters ($\chi^2/d.o.f.$ = 637.81/667). The best fit parameters for inclination, inner radius and emissivity index are: $\theta= 17_{-3}^{+2}$, $r_{in}<74\ r_g$ and $ q = 1.69_{-0.69}^{+0.48}$. The equivalent width of the relativistic line is $EW = 81_{-24}^{+27}$ eV. Our results for this source are in agreement within the errors with previous analysis of the same spectrum performed by \citet{Patrick+12}. Figure \ref{7469} shows the data to model ratios for NGC 7469 when the Baseline model (upper panel), the Narrow Line Model (middle panel) and the Relativistic Relline model (bottom panel) are applied to the data. It is clear that, when the inclination, inner radius and emissivity index are free to vary in the fit, the Relativistic Relline model is fitting all the residuals associated with the broad Fe component.

\begin{figure*}
\centering
\includegraphics[clip=true,width=0.75\textwidth]{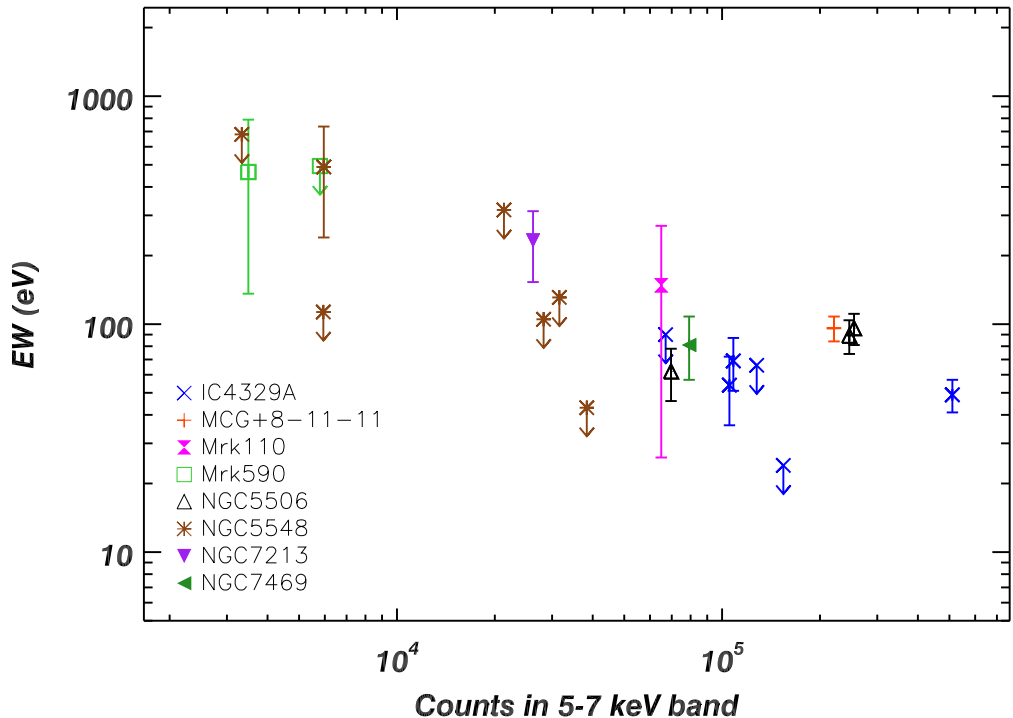}
\caption{Equivalent Width (eV) of the relativistic Fe K$\alpha$ line component as a function of the counts in the Fe energy band. These counts are calculated for each observation as the product of the flux in the 5-7 keV energy band and the exposure time.}
\label{EW}
\end{figure*} 

Fig. \ref{EW} shows the equivalent width of the relativistic Fe K$\alpha$ line for each observations as a function of the counts in the Fe K band. The latter values were calculated as the product between the counts/s in the 5-7 keV energy band and the exposure time of the observation. Different symbols and colours in the plot correspond to different sources as described in the legend. It is clear from the figure that when the observations have a low number of counts the measurement of the equivalent width of the relativistic line in general gives only an upper limit. In contrast, whenever the counts in the Fe band are higher than $\sim$ 4 $\times$ 10$^4$, we generally see evidence for the presence of a broad Fe K$\alpha$ line. Nevertheless, we do detect a relativistic Fe line in spectrum with fewer counts ($\sim$ 3 $\times$ 10$^3$). In Mrk 590, the inferred equivalent width of the line ($\sim$ 450 eV) is much larger than is typical, accounting for the detection. In 6 out of 7 observations of NGC 5548, we measured only an upper limit to the flux of the relativistic Fe line. In one observation, however, we find an apparently intense broad line with EW$\sim 500\pm250$ eV, albeit at low significance. It seems likely that this is due to Malmquist bias, whereby we see a positive statistical fluctuation in the line strength resulting in an apparent detection of a strong line, when the true EW is much lower. We further note that at some epochs exhibits absorption feature in the Fe K-band, which may complicated the detection of the broad Fe K emission (e.g. \citealt{Liu+10}, \citealt{Kaastra+14}). We note that while the trend of positive detections at high signal-to-noise ratio is clear, there are also spectra with high signal-to-noise ratio where only upper limits are obtained, most notably for IC4329A, where one such upper limit is inconsistent with the detections in the same source. This is discussed further in section \ref{SpecialCases}. The results on this source are consistent with the previous work of \citet{Mantovani+14} where the same \textit{Suzaku} observations were analyzed. The authors pointed out that the relativistic line in this object is relatively weak and detected with high significance only when the data are combined together. 

The main result of our analysis i.e. the need for high signal-to-noise ratio in the detection of broad lines, is in agreement with previous work using samples of sources observed with \textit{XMM-Newton} (e.g., \citealt{Guainazzi+06}, \citealt{Nandra+07}, \citealt{Bhayani+11}, \citealt{Calle+10}). 

\section{A Self-Consistent Reflection Model}

The Relativistic Relline model described above provides an improvement to the fit in many cases, but the model is not self-consistent. This is because, as the disk is expected to be optically thick, the relativistic line should be accompanied by a reflection continuum which is also modified by the same relativistic effects as the line. Furthermore, the strength of the continuum should be linked to the line. This latter consideration also applies to the reflection continuum associated with the narrow emission line, which was fitted separately from the line in our Baseline model. 
 
To provide a more physical and self-consistent model for the spectra, we used the \textit{pexmon} model (\citealt{Nandra+07}). This model reproduces the main features of the reflection spectrum produced by the interaction of the primary continuum with a Compton thick layer of neutral material (accretion disk and/or torus). This model combines: i) a narrow Fe K$\alpha$ line at 6.4 keV; ii) a narrow Fe K$\beta$ line at 7.06 keV; iii) a narrow Ni K$\alpha$ line at 7.47 keV; iv) Compton reflection and v) an Fe K$\alpha$ Compton shoulder. We adopt two such components, one from distant material (e.g. the torus) and another modified by relativistic effects in the accretion disk. We designate this the Relativistic Pexmon model (RP) \textit{zwabs*(cutoffpl+pexmon+relconv*pexmon)}. This model self consistently reproduces the high energy reflection continua associated with both the narrow and relativistic lines. To model the relativistic effects, one \textit{pexmon} component is convolved with the \textit{relconv} model (\citealt{Dauser+10}), which broadens the whole reflection spectrum using the same kernel as the \textit{relline} emission line model (\citealt{Dauser+10}). As in the Relativistic Relline model, the \textit{zwabs} model was introduced only for IC 4329A and NGC 5506 with column densities of N$_H$ = 0.4 $\times$ 10$^{22}$ cm$^{-2}$ and N$_H$ = 3 $\times$ 10$^{22}$ cm$^{-2}$, respectively. We also fixed the high energy cut-off for these two sources to 180 keV and 130 keV, while for the other sources in the sample we adopted 300 keV. The Fe abundance was always fixed to the solar value, and in the \textit{relconv} model we fixed the following parameters, as for the \textit{relline} model: i) index1 = index2 = 3; ii) R$_{break}$ = 15 r$_g$; iii) R$_{out}$ = 400 r$_g$; iv) limb = 0. For each source, we adopted the combination of inclination and inner radius as in the Relativistic Relline model (see Table \ref{relline}). The spin parameter was assumed to be a = 0 for r$_{in}$ = 6 r$_g$ and a = 0.998 for r$_{in}$ = 1.24 r$_g$. 

\begin{table*}
\caption{The table shows the $\chi^2$/d.o.f. for the Baseline model plus the relativistic Fe K$\alpha$ line (Relativistic Relline model) and for the self-consistent Relativistic Pexmon model. We also report the $\Delta\chi^2$/$\Delta$d.o.f. between the different models, expressed as $\chi^2_{RP}$-$\chi^2_{RR}$.}
\centering
\label{DeltaChi}
\resizebox{15cm}{!} {
\begin{tabular}{@{}ccccc@{}}
\hline
Source& Observation ID & Relativistic Relline Model & Relativistic Pexmon Model & $\Delta\chi^2$/$\Delta$d.o.f.\\
\hline
NGC 5506 & 701030010 & 1348.65/1334 & 1369.46/1335 & 20.81/1\\
& 701030020 & 1423.86/1403 & 1431.75/1404 & 7.89/1\\
& 701030030 & 1381.03/1377 & 1386.33/1378 & 5.3/1\\ \\
IC 4329A & 702113010 & 923.58/876 & 917.12/877 & -6.46/1\\
& 702113020 & 1092.91/1093 & 1148.52/1094 & 55.61/1\\
& 702113030 & 990.03/989 & 988.59/990 & -1.44/1\\
& 702113040 & 908.35/861 & 908.75/862 & 0.4/1\\
& 702113050 & 585.23/586 & 592.92/587 & 7.69/1\\
& 707025010 & 1656.27/1639 & 1658.91/1640 & 2.64/1\\ \\
MCG +8-11-11 & 702112010 & 1236.05/1324 & 1358.79/1325 & 122.74/1\\ \\
NGC 7213 & 701029010 & 671.73/689 & 672.52/690 & 0.79/1\\ \\
MRK 110 & 702124010 & 545.66/546 & 549.12/547 & 3.46/1\\ \\
NGC 7469 & 703028010 & 637.81/667 & 625.46/668 & -12.35/1\\ \\
NGC 5548 & 702042010 & 83.01/101 & 84.14/102 & 1.13/1\\
& 702042020 & 233.10/188 & 229.29/189 & -3.81/1\\
& 702042040 & 328.70/312 & 326.69/313 & -2.01/1\\
& 702042050 & 222.19/230 &  222.01/231 & -0.17/1\\
& 702042060 & 389.54/374 & 390.92/375 & 1.38/1\\
& 702042070 & 253.00/295 & 255.41/ 296 & 2.41/1\\
& 702042080 & 172.86/180 & 172.70/181 & -0.16/1\\ \\
MRK 590 & 705043010 & 198.71/186 & 198.37/187 & -0.34/1\\
& 705043020 & 102.35/116 & 100.94/117 & -1.41/1\\ \hline
\end{tabular}}
\end{table*}

Table \ref{DeltaChi} presents the $\chi^2$/d.o.f. for the Relativistic Relline and for the Relativistic Pexmon model. We also report the $\Delta\chi^2$ per $\Delta$d.o.f. between these two, expressed as $\chi^2_{RP}$-$\chi^2_{RR}$. This is helpful to understand whether the Fe emission line strength and the Compton hump at higher energies are simultaneously well-fitted, since the \textit{pexmon} model links these two quantities while the \textit{rellline} model does not. We find in general that the physically self-consistent model provides a similar fit to the phenomenological one, showing consistency between the strength of the Fe K$\alpha$ line and the reflection continuum at high energies (e.g. Compton hump). However, in two cases (MCG +8-11-11 and IC 4329A Obs. ID 702113020) the $\Delta\chi^2$ between the Relativistic Pexmon and the Relativistic Relline models is large ($\Delta\chi^2>50$), indicating a much better fit if the line and continuum are decoupled. We now consider these cases in more detail. 

\subsection{The cases of MCG +8-11-11 and IC 4329A}
\label{SpecialCases}

We investigate here the two extreme cases where the Relativistic Relline model gives better fit to the data, with a $\Delta\chi^2>50$, compared to the more self-consistent Relativistic Pexmon model: MCG +8-11-11 and one observation of IC 4329A (ID: 702113020). 

The case of MCG +8-11-11 has already been noted by \citet{Bianchi+10}. They found a detection of relativistic Fe K$\alpha$ line, but without any reflection component at higher energies. The results of our analysis are fully consistent with this result (see Table \ref{relline}). When the Relativistic Pexmon model is applied to the data, it leads to an inadequate fit with $\Delta\chi$ = 122.74 compared to the Relativistic Relline model. The reason is that, in the former, the Compton hump is linked to the strength of the emission line, while in the latter they are independent. This can be discerned immediately from the parameters in Table \ref{relline}. The reflection fraction $R$ for MCG +8-11-11 is very small and well constrained $R = 0.12_{-0.05}^{+0.06}$. Conversely both the narrow and relativistic lines are well detected and have a combined equivalent width of 156 eV, corresponding to a total reflection fraction of $R\sim1$. 

\begin{figure}
\begin{minipage}[b]{0.9\linewidth}
\centering
\includegraphics[width=0.85\textwidth, angle=270]{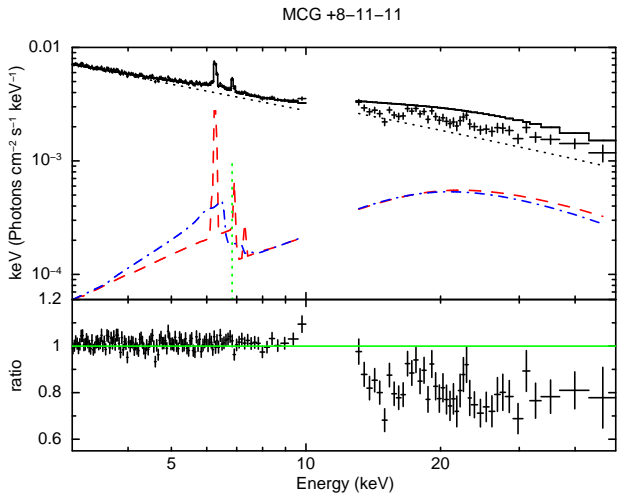}
\end{minipage}
\caption{Unfolded spectrum for MCG +8-11-11. The Relativistic Pexmon model is applied only to the XIS spectrum and extrapolated to the high energies. The bottom panel shows the data to model ratio. This model is overestimating the amount of reflection because of to the presence of the relativistic Fe K$\alpha$ line.}
\label{MCGextra}
\end{figure}

Figure \ref{MCGextra} shows the unfolded spectrum for this source together with all the components of the Relativistic Pexmon model. The plot shows the best fit model of the XIS spectrum extrapolated to the high energy band, and compared to the PIN data. The bottom panel shows the data to model ratio for both the XIS and PIN spectra. It is evident that the \textit{pexmon} model greatly overestimates the amount of reflection at high energies. 

Physically, there are several possible interpretations for the appearance of the spectrum of MCG +8-11-11. The first is that the line emission arises from material of small optical depth, producing a line but with little Compton scattering, and hence Compton hump. This interpretation is very unlikely: the line emission in this source is strong and low optical depth material would likely underproduce the emission line unless other circumstances were in play. A second possibility is that the material has very high Fe abundance. This could produce line emission which is relatively strong compared to the Compton hump. We tested for this possibility by allowing the Fe abundance to be a free parameter in the two \textit{pexmon} models. Letting this parameter free to vary, the fit does improve ($\chi^2$/d.o.f. = 1238.17/1323). However, the values of the Fe abundance for the two components of the \textit{pexmon} model are very extreme (A$_{Fe}$/A$_{Solar}$ $\sim$99). This explanation therefore also seems quite unlikely. While the Fe abundance can explain an anomalous ratio of the line to the continuum, it does not explain why the absolute value of the reflection fraction as measured by the Compton hump is so low in this source. The most likely explanation for the phenomenology of MCG +8-11-11 is related to our assumptions about the high energy cutoff. In our fit we have assumed an exponential cutoff of 300 keV. This ensures that there are plentiful high energy photons which can be downscattered into the Compton hump. If the temperature of the corona is lower, this will suppress the high energy reflection while having little effect on the strength of the Fe K$\alpha$ line. We tested this by fitting the spectrum imposing an high energy cutoff of E$_c$ = 50 keV in both the Relativistic Relline and the Relativistic Pexmon models. The $\chi^2$/d.o.f. now are comparable (Relativistic Relline: $\chi^2$/d.o.f. = 1262.03/1324; Relativistic Pexmon: $\chi^2$/d.o.f. = 1262.31/1325). 

Finally we tested the hypothesis of ionized reflection in this source. We fitted the spectrum with the \textit{relxill} model (\citealt{Garcia+14}) together with a narrow \textit{pexmon} component. The \textit{relxill} model allows us to fit the reflection spectrum produced by ionized material in the inner regions of the accretion disk, modified by relativistic effects. The best fit $\chi^2/d.o.f.$ for this model is 1254.91/1322, which is comparable to that obtained with the Relativistic Relline model. The model fit constraints the inclination, the reflection fraction from the disk and the high energy cut-off of the primary continuum: $\theta = 38_{-5}^{+4}$, $ R = 0.20_{-0.08}^{+0.10}$ and $E_c = 68_{-11}^{+15}$ keV. Ionization of the disk is thus also a plausible explanation for the unusual spectrum, although the rather low reflection fraction is puzzling.

\begin{figure}
\begin{minipage}[b]{0.9\linewidth}
\centering
\includegraphics[width=0.85\textwidth, angle=270]{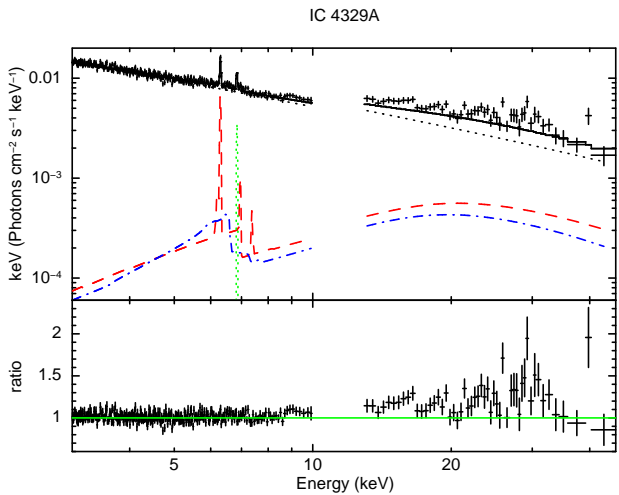}
\end{minipage}
\caption{Unfolded spectrum for IC 4329A (Obs. 702113020). The Relativistic Pexmon model is applied only to the XIS spectrum and extrapolated to the energy range of the PIN data. The bottom panel shows the data to model ratio. In this observation the Fe K$\alpha$ line is not detected, however there is a significant reflection component. The \textit{pexmon} model constrained by the line underestimates the reflection at high energies.}
\label{ICextra}
\end{figure}

Intriguingly, the case of IC 4329A Obs. 702113020 appears to be opposite to that of MCG +8-11-11. In this observation, a relativistic Fe K$\alpha$ line is not detected. However, a significant contribution of the reflection continuum at high energies is present, and it is well constrained (see Table \ref{relline}). This is a particularly puzzling case, because the relativistic Fe K$\alpha$ feature, while relatively week, is detected in all the other \textit{SUZAKU} spectra of this object (see also \citealt{Mantovani+14}). Figure \ref{ICextra} shows the unfolded spectrum for this observation when the Relativistic Pexmon model is applied only to the XIS spectrum and extrapolated to high energies, and compared to the PIN data. 

Formally a sub-solar Fe abundance can account for the weak Fe K$\alpha$ line in this individual observation. We let the Fe abundance vary in both the Relativistic Relline and Relativistic Pexmon models. The best-fit Fe abundance for the two models indeed appear to be sub-solar (Relativistic Relline: 0.46$_{-015}^{+0.21}$; Relativistic Pexmon: 0.25$_{-0.08}^{+0.09}$). The difference in the goodness of fit between the two also reduces substantially, with $\Delta\chi^{2} =10.67$, much less then the the $\Delta\chi^2$ = 55.61 obtained with solar abundances. However, the other observations of IC 4329A have Fe abundance close to solar. A variation of the Fe abundances is highly unlikely, especially if we consider the short time scale between each observation ($\sim$ 1 week). These observations of IC 4329A therefore provide a challenge to the prevailing reflection paradigm.  

\section{Discussion}

We have analysed a sample of Seyfert 1 galaxies observed with \textit{Suzaku}, which had previously shown an absence of evidence for a relativistic Fe K$\alpha$ emission line in their \textit{XMM-Newton} spectra. The goals of this work were to investigate whether the \textit{Suzaku} spectra show evidence for the relativistic line, and whether the line, if present, is consistent with the measured strength of the Compton hump. To achieve these aims, we systematically fitted the spectra of all the sources with a series of models of varying complexity, testing for the presence of relativistic lines and their consistency with the Compton reflection hump at higher energies.

As in previous work, we find that all the spectra show a peak at 6.4 keV which we identify with a narrow line from relatively distant material. The majority of the sources show additional complexity around the Fe K-complex, however. Several sources also show evidence for more highly ionized species, identified with Fe {\sc xxv} and Fe {\sc xxvi}. In addition, at least one observation of all the sources in our sample, and 12 out of 22 in total, shows a significant improvement when a relativistic line is added to the model. Where broad lines are detect in our sample, the typical equivalent widths are in the range 50-100 eV, and the upper limits in the case of non-detections rarely exclude such values. This is fully consistent with the work of \citet{Nandra+07}, using XMM-Newton data, who found an average broad line EW of $\sim 80$ eV. Both are consistent with the work of \citet{Patrick+12} where a sample of Seyfert 1 AGN was analyzed using \textit{Suzaku} observations. In this case, the mean EW within the sample was found to be 96$_{-10}^{+10}$ eV. It is interesting to note that the sources in our sample were chosen because the relativistic line in the previous \textit{XMM-Newton} spectra is weak and/or of low significance, but our results show that in reality they are entirely consistent with typical line strength seen in more complete samples.

The results for the individual sources in our sample are also broadly consistent with previous work. \textit{XMM-Newton} observations of NGC 5506 revealed the presence of a broad residual in the Fe band (\citealt{Matt+01}). The overall emission complex was equally well fit either with a broad line or a blend of He-like and H-like narrow lines. Our analysis of this source shows that, even accounting for the presence of the ionised emission lines, a relativistic Fe K$\alpha$ emission component with an equivalent width of $\sim$82 eV is required in addition. The same observations were analyzed by \citet{Patrick+12} where a relativistic line was detected but with a lower value of equivalent width ($\sim$ 30 eV). In this work the parameters of the \textit{relline} component, such as the emissivity index, the inclination and the inner radius, were left free to vary, whereas some parameters were fixed in our work. This may partially account for the difference in derived broad line EW. Recently, a \textit{NuSTAR} observation of NGC 5506 was analyzed by \citet{Matt+15}. A significant relativistic Fe K$\alpha$ line was not required in their fitting of that spectrum, perhaps suggesting a variation of the intensity of the line between the \textit{Suzaku} and \textit{NuSTAR} observations, which are separated by $\sim$7 years.

The results of the analysis on the bright Seyfert 1 galaxy, IC 4329A are consistent with the work of \citet{Mantovani+14}, where the same five \textit{Suzaku} observations were analysed, plus an additional Suzaku observation which is simultaneous with a \textit{NuSTAR} observation. This spectrum was analysed by \citet{Brenneman+14} but only a moderately broad Fe K$\alpha$ line was detected using a simple gaussian model with an equivalent width of 34$_{-7}^{+8}$ eV. Considering only one observation of this source, it was not possible to constrain any parameters of the relativistic model for the Fe K$\alpha$ line (\citealt{Brenneman+14}). In this work, we considered all 6 \textit{Suzaku} observations available in archive, which allowed us to better constrain the parameters of the relativistic line, such as the inclination of the disk (30$^{\circ}$) and the inner radius (6 r$_g$), which are also in agreement with the previous work of \citet{Mantovani+14}. We detected the relativistic line at 95 per cent confidence in 3 observations out of 6, with a mean value for the equivalent width of $\sim$ 86 eV. \citet{Mantovani+14} pointed out that this spectral feature was detected only in the combined data because of the weakness of the feature and relatively poor statistics of the single observations. Our results on this source are fully consistent with this scenario. The same observations were analyzed by \citet{Patrick+12}. Using the same model component for modelling the relativistic Fe line (i.e. \textit{relline}), and they found an equivalent width of 69$_{-14}^{+13}$ eV, slightly lower than the value found in our analysis. This difference could be due to the fact that they modelled the mean spectrum and not the single observations separately and also again to the fact that the parameters of the \textit{relline} model were free to vary.

As discussed above, a relativistic Fe K$\alpha$ line was already detected in MCG +8-11-11 in the same \textit{SUZAKU} observation with an equivalent width of about 90 eV (\citealt{Bianchi+10}), and our measurements are consistent with these results (EW = 92$_{-23}^{+23}$ eV).

Previous work on NGC 7213 (\citealt{Lobban+10}) has not revealed the presence of a relativistic emission line, whereas our analysis shows this component to be present in the \textit{Suzaku} spectra. In the work of \citet{Lobban+10} the same \textit{Suzaku} observation of NGC 7213 was considered. In their work, a \textit{diskline} model was used assuming an inclination of the disk of 30$^{\circ}$ and the inner radius of 6r$_g$, but an upper limit was found for the flux of the relativistic Fe line. However, in our work different combinations of inner radii and inclinations were tested. We do detect a relativistic Fe K$\alpha$ line with an equivalent width of 233$_{-80}^{+80}$ eV adopting an inclination of 80$^{\circ}$ and inner radius of 6r$_g$.

As with the other objects in our sample, Mrk 110 is a source which did not show evidence for a relativistic Fe line in previous \textit{XMM-Newton} data (\citealt{Nandra+07}). We do find such evidence, albeit at marginal statistic significance ($\sim$95 per cent confidence). The same \textit{Suzaku} observation was also analyzed by \citet{Walton+13}, who also found evidence for a broad Fe component.

\textit{Beppo-SAX} observations of NGC 7469 revealed the presence of both the relativistic Fe K$\alpha$ line (EW = 121$_{-100}^{+100}$ eV) and the associated Compton hump (\citealt{DeRosa+02}). Our results on this source are consistent with this work (EW =  81$_{-24}^{+27}$ eV). The same \textit{Suzaku} spectrum was also analyzed by \citet{Patrick+12}, who fitted the data with a relativistic line with an equivalent width consistent within the errors with our result (EW= 91$_{-8}^{+9}$ eV).

Finally, there are two objects in our sample where the evidence for any  relativistic Fe K$\alpha$ emission line is very weak, those being NGC 5548 and Mrk 590. We note that no broad Fe K$\alpha$ line was detected also in previous observations of these objects (NGC 5548: \citealt{Brenneman+12}, Mrk 590: \citealt{Longinotti+07}). In both of these sources, however, the constraints on any broad component in the spectrum are poor, with upper limits typically several 100 eV, usually in excess of the typical EW for the detected broad lines of $\sim50-100$ eV. The apparent absence of a broad features is therefore like due to the low signal-to-noise ratio of the data. 

Overall, our results are broadly consistent with the idea that relativistic Fe K$\alpha$ emission line is ubiquitous in the X-ray spectra of Seyfert 1 galaxies. The lack of detection in some sources and/or observations can be attributed to the low signal-to-noise ratio of the spectra analyzed. This result confirms previous conclusions from the analysis of \textit{XMM-Newton} spectra, where it was claimed that high signal-to-noise ratio is necessary for the detection of broad Fe lines of typical equivalent width (e.g. \citealt{Guainazzi+06}; \citealt{Nandra+07}; \citealt{Calle+10}). Our sample analysis shows that broad line are consistently detected when the counts in the 5-7 keV energy band are $\gtrsim$ 4 $\times$ 10$^4$.

Taking advantage of the extended energy band of \textit{SUZAKU}, we were also able to examine the relationship between the emission of the Fe K$\alpha$ line and the reflection continuum at higher energies. In particular, we compared the fits with our Relativistic Relline model, which allows the Fe line and reflection strengths to vary independently, with a more self-consistent reflection model (Relativistic Pexmon model), in which the line and Compton hump are linked in the ratio expected for a Compton-thick slab. For the vast majority of the observations (19/22), this self-consistent model gives a better or comparable fit to the phenomenological Relativistic Relline model. This shows that in general the line and reflection strengths in AGN are consistent with each other, in support of the idea that both arise from the same material. In two cases the self-consistent model leads to an inadequate fit, that of MCG +8-11-11 and in one of the spectra of IC 4329A, which shows a difference in fits statistic of $\Delta\chi^2$ $>$ 50. Surprisingly, these two observations exhibit this behaviour for opposite reasons. 

In MCG +8-11-11, as already noted by \citet{Bianchi+10}, strong evidence for a relativistic Fe line is present, but with the absence of any reflection emission at high energy. A possible explanation for the lack of reflection in this source could be a low value of the high energy cut off (e.g. $\sim$ 50 keV). This hypothesis could be tested using high energy data of higher quality, as can be expected e.g. from \textit{NuSTAR}. 

The opposite situation pertains to one observation of IC 4329A (Obs. ID: 702113020). In this spectrum the relativistic line was not detected, with a tight upper limit, while a strong reflection continuum is observed at high energies. This is very puzzling, especially when considered in context with the other observations of IC 4329A with \textit{Suzaku}. These do show both a relativistic line and a reflection continuum, with strengths that are typically consistent with each other. The lack of consistency in just one case argues against some interpretations, such as an unusual Fe abundance in the reflecting material. A possible explanation for the results is rapid variability of the geometry of the inner regions of the disk, for example variability of the illuminating pattern and/or ionization state of the disk, or strong relativistic effects (e.g. \citealt{Iwasawa+96}; \citealt{Miniutti+03}). These might results in unusual line and continuum properties, such as strong line profile variations, or extreme broadening of the line, which may make it difficult to distinguish from the continuum. Overall, however, it is difficult to provide a robust interpretation for why in this source the line and reflection continuum would not follow each other. If the two components can exhibit differential variability of this kind, it is not only difficult to explain, but also difficult to make predictions which would allow models to be tested. This is troublesome for the standard reflection models, and unless an interpretational framework can be established, diminishes their diagnostic power.


\section*{Acknowledgments}

This research has made use of data obtained from the Suzaku satellite, a collaborative mission between the space agencies of Japan (JAXA) and the USA (NASA). G.P. acknowledges support via an EU Marie Curie Intra- European fellowship under contract no. FP-PEOPLE-2012-IEF- 331095 and the Bundesministerium f{\" u}r Wirtschaft und Technologie/Deutsches Zentrum f{\" u}r Luft-und Raumfahrt (BMWI/DLR, FKZ 50 OR 1408) and the Max Planck Society. We thank the anonymous referee for his/her constructive comments. GM acknowledges M. Fossati, B. De Marco and A. Ballone for useful discussions and suggestions.

\bsp

\label{lastpage}

\end{document}